\documentclass[journal=jacsat,manuscript=article]{achemso}

\usepackage[version=3]{mhchem} 
\usepackage[utf8]{inputenc}
\usepackage{graphicx}
\usepackage{dcolumn}
\usepackage{bm}
\usepackage{blindtext}
\usepackage{fancyvrb}
\usepackage[normalem]{ulem}

\usepackage[justification=centering]{caption}
\newcommand\Tstrut{\rule{0pt}{2.9ex}}         
\newcommand\Bstrut{\rule[-1.2ex]{0pt}{0pt}}   
\newcommand\TBstrut{\Tstrut\Bstrut}           
\newcommand{\ttiny}[1]{\text{\tiny{#1}}}

\usepackage{tikz}
\usepackage{ulem}
\usetikzlibrary{shapes,arrows}
\usepackage{multirow}
\usepackage{amsmath}

\usepackage{siunitx,letltxmacro}
\LetLtxMacro{\svqty}{\qty}
\usepackage{physics}
\LetLtxMacro{\qty}{\svqty}
\author{Nastasia Mauger}
\affiliation{Sorbonne Université, Laboratoire de Chimie Théorique, UMR 7616 CNRS, 75005, Paris, France}
\author{Thomas Plé}
\affiliation{Sorbonne Université, Laboratoire de Chimie Théorique, UMR 7616 CNRS, 75005, Paris, France}
\author{Louis Lagardère*}
\affiliation{Sorbonne Université, Laboratoire de Chimie Théorique, UMR 7616 CNRS, 75005, Paris, France}
\author{Simon Huppert*}
\affiliation{Sorbonne Universit\'e, Institut des NanoSciences de Paris,
UMR 7588 CNRS, 75005, Paris, France}
\author{Jean-Philip Piquemal*}
\affiliation{Sorbonne Université, Laboratoire de Chimie Théorique, UMR 7616 CNRS, 75005, Paris, France}
\email{louis.lagardere@sorbonne-universite.fr, simon.huppert@sorbonne-universite.fr, jean-philip.piquemal@sorbonne-université.fr}

\title{Improving Condensed Phase Water Dynamics with Explicit Nuclear Quantum  Effects: the Polarizable Q-AMOEBA Force Field}

\abbreviations{IR,NMR,UV}

\begin{document}


\begin{abstract}
We introduce a new parametrization of the AMOEBA polarizable force field for water denoted Q-AMOEBA, for use in simulations that explicitly account for nuclear quantum effects (NQEs). This study is made possible thanks to the recently introduced adaptive Quantum Thermal Bath (adQTB) simulation technique which computational cost is comparable to classical molecular dynamics. The flexible Q-AMOEBA model conserves the initial AMOEBA functional form, with an intermolecular potential including an atomic multipole description of electrostatic interactions (up to quadrupole), a polarization contribution based on the Thole interaction model and a buffered 14-7 potential to model van der Waals interactions. It has been obtained by using a Force Balance fitting strategy including high-level quantum chemistry reference energies and selected condensed phase properties targets. The final Q-AMOEBA model is shown to accurately reproduce both gas phase and condensed phase properties, notably improving the original AMOEBA water model. This development allows the fine study of NQEs on water liquid phase properties such as the average H-O-H angle compared to its gas phase equilibrium value, isotope effects etc... Q-AMOEBA also provides improved infrared spectroscopy prediction capabilities compared to AMOEBA03. Overall, we show that the impact of NQEs depends on the underlying model functional form and on the associated strength of hydrogen bonds. Since adQTB simulations can be performed at near classical computational cost using the Tinker-HP package, Q-AMOEBA can be extended to organic molecules, proteins and nucleic acids opening the possibility for the large scale study of the importance of NQEs in biophysics.
\end{abstract}

\section{Introduction}
Classical molecular dynamics (MD) has become a powerful valuable tool to study the properties of complex systems, such as condensed matter, organic molecules and proteins, but also to design new molecules and drugs \cite{goh2016computational,misquitta2014distributed,walsh2017biointerface,Durrant2011,Karplus2002,HANSSON2002190}. It provides access to most of the relevant thermodynamic observables of a given system. The quality of these observables relies on the accuracy of the potential energy surface (PES), usually called Force Field (FF), but also on the amount of the phase-space sampling.
Thanks to the different advances in computing hardware, such as GPU computing \cite{kutzner2015best,eastman2017openmm,salomon2013routine,Adjoua2021}, and techniques such as enhanced sampling \cite{elber2016perspective,faradjian2004computing},  MD is now capable of reaching timescales of ms making it possible the extensive study of macromolecules \cite{Pierce2012}.

Most standard additive FFs, such as AMBER \cite{Case2005-xl,https://doi.org/10.1002/wcms.1121}, CHARMM \cite{https://doi.org/10.1002/jcc.21287,MacKerell1998,Lopes2015-qd,Vanommeslaeghe2010-dq}, OPLS \cite{Jorgensen1988,Jorgensen1996,Kaminski2001}, COMPASS \cite{Sun1998}, have been widely used in MD thanks to their low computational cost due to their relatively simple functional form and they have been refined throughout the years. However, they lack a proper description of many body interactions and, although they provide good results for various properties compared to experiments, it appears that their accuracy is limited in other contexts \cite{10.3389/fmolb.2019.00143}. The major limitation of these FFs is the use of fixed partial atomic charges to model electrostatic interactions and their lack of a fine description of many-body polarization which hinders their transferability when the atomic environment changes. To circumvent that, much effort has been made to include explicitly many-body polarization in FFs by using either Drude oscillators, fluctuating charges or induced dipoles.\cite{10.3389/fmolb.2019.00143,chapterpol,doi:10.1146/annurev-biophys-070317-033349} Versions of CHARMM \cite{Patel2004-hp,Patel2004-qr} and AMBER \cite{https://doi.org/10.1002/jcc.20386} have been published where polarization was added to the existing non polarizable FF and new ones have been developed such as PFF \cite{Kaminski2002-bi}. Moreover, it has been shown that FFs should include the polarization from scratch \cite{Mei2015}. Hence, more modern FFs such as AMOEBA \cite{Ren2003,Ponder2010}, AMOEBA+ \cite{AMOEBA+,AMOEBA+2} or SIBFA \cite{Naseem-Khan2022,NohadGresh2007} have been designed with polarization included on the outset. 

Due to its major role in many fields, water has been a central focus of FF development. Over the past decades, many FFs have been developed, such as TIP3P~\cite{doi:10.1063/1.445869}, SPC~\cite{berendsen1981interaction} and their variants~\cite{doi:10.1063/1.3394863,doi:10.1063/1.481505}. Although widely used, these models struggle to describe a wide part of the phase diagram of water and ice. For example, the gas phase binding of the water dimer is overestimated by approximately 30$\%$ in the TIP5P model~\cite{doi:10.1063/1.481505}. Based on this, other water models have been developed such has BK3~\cite{kiss2013systematic}, TIP4P/Ew~\cite{doi:10.1063/1.1683075} and TIP4P/2005~\cite{B805531A}, which uses implicit many-body effects through classical polarization to go beyond the pairwise approximation. However, these models were parametrized for classical simulations and do not take into account explicitly Nuclear Quantum Effects (NQEs) which reduce their transferability. Indeed, such effects are important for systems involving light atoms such as hydrogen~\cite{Ceriotti2016} and some studies have highlighted their importance in biological systems~\cite{agarwal2002nuclear,zhang2022quantum}. In practice, such effects are usually implicitly added by fitting the analytical PES to recover thermodynamic observables. Moreover, many recent water models use high quality \textit{ab initio} data as a basis for the PES and it has been shown that with such a high precision, NQEs must be taken into account explicitly to accurately reproduce thermodynamic observables~\cite{doi:10.1073/pnas.1806064115}.

A rigorous and asymptotically exact approach to include NQEs explicitly in MD simulations is provided by the path integral molecular dynamics (PIMD) formalism where the dynamics is performed on an extended classical system consisting in several replicas (also called beads) of the physical system \cite{feynman2010quantum, chandlerJCP1981PI}. This method allows for a systematically improvable treatment of NQEs, but at a computational cost that is considerably larger than that of classical MD. This increased cost has so far limited the development of FF models based on PIMD estimations of thermodynamic properties, with a few notable exceptions such as TIP4PQ/2005 \cite{doi:10.1063/1.3175694}, q-SPC/Fw \cite{doi:10.1063/1.2386157}, q-TIP4P/F \cite{doi:10.1063/1.3167790}, and more recently ArrowFF \cite{doi:10.1073/pnas.1806064115}. 
In parallel, different methods have been developed to reduce the computational cost of PIMD simulations, such as high-order PI~\cite{doi:10.1063/1.4971438,doi:10.1063/1.3609120}, PI perturbation theory~\cite{Poltavsky2020,C5SC03443D} 
or path integral generalized Langevin methods~\cite{brieuc_JCTC2016_QTB_PI, ceriotti_JCP2011_PIGLET}. Though these methods have enabled significant progress, PIMD simulations remain significantly costlier than classical MD (typically requiring one order of magnitude more resources at room temperature). Other approaches, such as ring-polymer contractions~\cite{markland2008efficient,markland2008refined} or imaginary multiple time-stepping~ \cite{cheng2016accelerating} allow reducing the number of replicas used in the evaluation of long-range interaction terms, thereby achieving almost classical computational cost in some applications~\cite{fanourgakis2009fast}. However the splitting of the PES between long-range and short-range terms is not always available, in particular in the context of \textit{ab initio} MD or when using \textit{machine learning} potentials. 
In this paper, we introduce a new polarizable force field, Q-AMOEBA, based on the AMOEBA functional form that is designed to be used with the explicit inclusion of NQEs. Indeed, since experiments naturally contain “quantumness”, the initial AMOEBA model intended to implicitly include quantum nuclear effects. When adding explicitly NQEs, one needs to reparametrize the force field potential to propose a version of it that only tends to reproduce the Born Oppenheimer energy in order to avoid double counting when NQEs are explicitly included through the dynamics. To do so, we make extensive use of the quantum thermal bath (QTB) a method based on a generalized Langevin thermostat to approximate the zero point motion of the nuclei. More precisely, we use the adaptive QTB method (adQTB) that relies on the quantum fluctuation-dissipation theorem to systematically compensate the effects of zero-point energy leakage (ZPEL)~\cite{Mangaud2019}. This approach was recently shown to reliably approximate NQEs in liquid water at a computational cost that remains similar to that of classical MD~\cite{doi:10.1021/acs.jpclett.1c01722}. 
The Q-AMOEBA force field parameters are optimized with the Force Balance software (FB)~\cite{Wang2013} in combination with Tinker-HP~\cite{C7SC04531J,Adjoua2021} package for molecular dynamics simulations, in order to accurately reproduce energies of various water systems in the gas phase as well as a few condensed phase properties obtained with a quantum description of the nuclei. This model will then make the extensive description of more complex systems involving water with explicit NQEs possible. Q-AMOEBA gives overall better results on energies and on key observables such as the isobaric heat capacity or the thermal expansion coefficient than the original AMOEBA model. Such a model also allows to finely study the practical effects of NQEs by comparing the properties that are obtained with it but with purely classical dynamics.
We first describe our parametrization strategy in details and then test the accuracy and transferability of our model on various thermodynamic and structural properties as well as on the IR spectra, showing the validity of our approach and its future applicability to the development of new generation models.
\section{Methods}
\subsection{AMOEBA Model}
The total potential energy of the AMOEBA \cite{ren2003polarizable} water model can be expressed as the sum of bonded and non-bonded energy terms:
\begin{equation}
    \begin{split}
   & E_{total}=E_{bonded}+E_{nonbonded} \\
   & E_{bonded}=E_{bond}+ E_{angle}+E_{b\theta} \\
   & E_{nonbonded}=E_{vdW}+E_{ele}^{perm}+E_{ele}^{ind}
    \end{split}
\end{equation}
The bonded terms which include the anharmonic bond-stretching and angle-bending terms are similar to MM3 force field~\cite{allinger1989molecular}. The water intramolecular geometry and vibrations are described with an Urey-Bradley terms $E_{b\theta}$. In the original AMOEBA model~\cite{Ren2003}, the ideal bond length was chosen to be at the experimental value of 0.9572 \AA, the ideal bending angle to 108.5$^{\circ}$ and the Urey-Bradley ideal distance was set to 1.5326 \AA. The angle is larger than the experimental gas-phase angle of 104.52$^{\circ}$ which was shown to be necessary to reproduce the correct average experimental angle in liquid water~\cite{ren2003polarizable}.

The non-bonded terms are composed of the vdW interactions and the electrostatic contributions from both permanent and induced dipoles. 
The vdW functional term uses Halgren's buffered 14-7 potential to model the pairwise additive interactions for dispersion at long-range and exchange-repulsion at short-range \cite{halgren1992representation}:
\begin{equation}
    E_{vdW}=\varepsilon_{ij}\left(\frac{1+\delta}{\sigma_{ij}+\delta}\right)^7\left(\frac{1+\gamma}{\sigma_{ij}^7+\gamma}-2\right)
\end{equation}
with $\varepsilon_{ij}$ the potential well depth and $\sigma_{ij}=r_{ij}/r_{ij}^0$ where $r_{ij}$ is the \textit{i}-\textit{j} separation and $r_{ij}^0$ is the minimum energy distance (distances are calculated from the oxygen atoms, see \cite{Ren2003}) . In AMOEBA, the vdW parameters were set to $\gamma=0.12$ and $\delta=0.07$.
The combining rules used for $r_{ij}^0$ and $\varepsilon_{ij}$ are given by:
\begin{equation}
    \begin{split}
       & r_{ij}^0=\frac{\left ( r_{ii}^0 \right)^3+\left ( r_{jj}^0 \right)^3}{\left ( r_{ii}^0 \right)^2+\left ( r_{jj}^0 \right)^2} \\
       & \varepsilon_{ij}=\frac{4\varepsilon_{ii}\varepsilon_{jj}}{\left (\sqrt{\varepsilon_{ii}}+\sqrt{\varepsilon_{jj}}\right)^2}
    \end{split}
\end{equation}
A hydrogen reduction factor is added which moves the hydrogen vdW center towards the oxygen along the O--H bond. 
To compute the electrostatic interactions the AMOEBA model uses point atomic multipoles truncated at quadrupoles for each atom center. The atomic multipole moments are derived using the distributed multipole analysis (DMA) approach and then optimized against a high-level ab-initio PES \cite{stone2005distributed}. More details about the functional form of AMOEBA can be found in reference \cite{Ponder2010}.
\subsection{Parametrization of Q-AMOEBA}
   
The Q-AMOEBA water model has been parametrized towards liquid phase simulations using the ForceBalance (FB) software \cite{wang2014forcebalance,wang2013systematic,wang2014building}. FB calculates the derivatives of the target properties with respect to the FF parameters to be optimized (see Appendix \ref{app:fb_deriv} for more details) and uses a Newton-Raphson procedure to reach a minimum with respect to a given objective function. In this work, the target properties include both \textit{ab initio} configurational energies and experimental observables. The initial parameters were taken from the original AMOEBA water model \cite{Ren2003} (denoted as AMOEBA03 in the rest of the text) and the weights of the different data included in the objective function are summarized in Table \ref{datas FB}. The different convergence criteria used for FB are given in SI. 

Most parameters of AMOEBA03 (such as the atomic permanent multipolar moments) are obtained directly from \textit{ab initio} calculations, therefore only van der Waals parameters and some well-chosen intramolecular terms were modified. More specifically, the optimization was performed on both van der Waals radii and epsilon values of the Halgren 14-7 potential \cite{halgren1992representation} associated to oxygen and hydrogen atoms, as well as the buffer radius of the hydrogen atom and equilibrium distances and angles of the intramolecular stretching and bending terms. 
Our initial intent was to develop a single set of parameters to be used with both PIMD and adQTB approaches. In practice we observed that separate FB optimizations with both methods yielded slightly different results. In particular, this is due to the particularities of pressure estimation in adQTB simulations (see Appendix \ref{app:pressure}), which leads to small inaccuracies in the determination of the density with this method (of the order of 1\%, as was already noted in a former study on a different FF model \cite{doi:10.1021/acs.jpclett.1c01722}). Considering the importance of this observable, we resorted to two sets of parameters with specific optimization for each method: one for PIMD denoted (1) and a second for adQTB, denoted (2).
Because the adQTB density appeared systematically slightly underestimated, we decided to further modify the buffer radius of H and the epsilon of O as they displayed the most significant discrepancies compared to the PIMD ones. To do so, we used the PIMD values as a guess and further used FB to finally obtain a satisfactory density with adQTB. This strategy ultimately led to the final adQTB parameters discussed in the rest of the paper. We emphasize that the two final sets of Q-AMOEBA parameters differ only slightly from each other.  
To perform the simulations of the condensed phase properties with FB, we used the Tinker-HP software on GPUs for PIMD and adQTB methods.

Geometries extracted from condensed-phase simulations while using the original AMOEBA03 model at temperatures ranging from 249.15K to 373.15K and including cluster sizes from 2 to 22 molecules were used to obtain ab initio references. The calculations were performed using Q-Chem 4.0 and include both energies and gradients calculated using the aug-cc-pVTZ basis set. Optimal geometries and binding energies of 40 small water clusters ranging from 2 to 20 molecules were also used. Each cluster has been used on the highest level of theory available, see table~\ref{datas FB} for details. 

The objective function includes two thermodynamic properties with experimental reference: the density $\rho$ and the enthalpy of vaporization $\Delta H_{vap}$. These properties were evaluated at different temperatures ranging from 249.15K to 369.15K under 1 atm. All simulations were carried out for \textit{N}=4000 molecules in  periodic boundary conditions with a cubic box. The van der Waals (vdW) cutoff was set to 12\AA\  and electrostatic interactions were computed with the SPME method \cite{essmann1995smooth} with a real space cutoff of 7\AA\ and a 60x60x60 grid. A RESPA integrator using a bonded/non-bonded split with timesteps of 0.2 and 2fs respectively  was used. 
For each FB iteration, the dynamics include 1 ns of equilibration 
and 4ns of production. For the PIMD calculations, 32 beads were used with the TRPMD (Thermostated Ring Polymer Molecular Dynamics) algorithm \cite{rossi2014remove} combined with a mild Langevin thermostat on the centroid (friction coefficient $\gamma=1\,\, \text{ps}^{-1}$), whereas in the adQTB method the friction was set to 20 ps$^{-1}$.
\begin{table}
\begin{tabular}{cccccc}
\hline \hline
System   & Reference data                          & Data type       & Data point & \multicolumn{2}{c}{Weight}               \TBstrut \\ \hline \hline
Clusters & Gas Phase Dipole-Quadrupole             & CCSD(T)         &            & 1.0                  & \multirow{6}{*}{} \TBstrut \\
         & Gas Phase Vibrational modes             & CCSD(T)         &            & 1.0                  &                  \TBstrut \\
         & Smith Dimer                             & CCSD(T) BE      & 10         & 1.0                  &                 \TBstrut  \\
         & Small Gas Phase Cluster                 & CCSD(T) BE      & 21         & 1.0                  &                \TBstrut   \\
         & Large Gas Phase Clusters                & MP2 BE      & 18         & 1.0                  &                  \TBstrut \\
         & PE and AF & MP2 & 42.000     & 1.0                  &                  \TBstrut \\
Liquid   & $\rho$                               & Expt.    & 10         &1.0 & 0.6             \TBstrut  \\
         & $\Delta H_{vap}$                & Expt.    & 10         &                      & 0.4               \TBstrut
\end{tabular}
\caption{\label{datas FB} Reference data used in FB to derive the Q-AMOEBA sets of parameters. BE refers to Binding Energies, PE to Potential Energies and AF to Atomistic Forces. 
The table also shows the weights that were applied to the different target properties. }
\end{table}

Because of the high frequencies of the intramolecular stretching and bending modes, the associated zero point energies are large, impacting the average values of the O-H distance and H-O-H angle. We therefore modified the corresponding parameters in Q-AMOEBA to recover experimental values. 

The final parameters of our models can be found in SI.

\section{Results and Discussion}

\subsection{Binding Energy of Water Clusters}
Tables \ref{BE of small Water Clusters} and \ref{BE of big Water Clusters} show the comparison of the configuration energies obtained with AMOEBA and Q-AMOEBA compared to accurate quantum references for water clusters of different sizes. For numbers of atoms $n<6$, the CCSD(T) level of theory has been used whereas larger clusters, such as octamers \cite{doi:10.1063/1.1626624}, 16-17 mers \cite{Yoo2010} and 20-mers \cite{doi:10.1063/1.1767519} were investigated at the MP2 level. These studies provide the optimized structures and binding energies (BEs) of these clusters that were used as a reference for the Q-AMOEBA parametrization. For a given model, the binding energy is defined as the difference between the optimized cluster energy and the sum of the energies of the optimized monomers.
 \begin{table*}
    \centering
    $
    \begin{array}{*{12}{c}}
    \hline \hline     
    \text{\small{(H$_2$O)$_2$}}     & \text{\small{CCSD(T)} }    &     \text{\small{AMOEBA03}}   & \text{\small{Q-AMOEBA (PIMD)}}  & \text{\small{Q-AMOEBA (adQTB)} } \TBstrut  \\
    \hline
    \text{\small{Smith01}}     & \text{\small{-4.97} }    &     \text{\small{-4.58}}   & \text{\small{-4,96}} & \text{\small{-4,95}}\TBstrut  \\
    \text{\small{Smith02}}     & \text{\small{-4.45} }    &     \text{\small{-3.98}}   & \text{\small{-4.35}}  & \text{\small{-4,34}} \TBstrut  \\
    \text{\small{Smith03}}     & \text{\small{-4.42} }    &     \text{\small{-3.94}}   & \text{\small{-4.3}}  & \text{\small{-4.3}}  \TBstrut  \\
    \text{\small{Smith04}}     & \text{\small{-4.25} }    &     \text{\small{-3.54}}   & \text{\small{-3.49}}  & \text{\small{-3.43
}}  \TBstrut  \\
    \text{\small{Smith05}}     & \text{\small{-4.00} }    &     \text{\small{-2.69}}   & \text{\small{-3.06}}  & \text{\small{-3.00}} \TBstrut  \\
    \text{\small{Smith06}}     & \text{\small{-3.96} }    &     \text{\small{-2.59}}   & \text{\small{-2,95}}  & \text{\small{-2.90}} \TBstrut  \\
    \text{\small{Smith07}}     & \text{\small{-3.26} }    &     \text{\small{-2.55}}   & \text{\small{-2,81}}  & \text{\small{-2.73}} \TBstrut  \\
    \text{\small{Smith08}}     & \text{\small{-1.30} }    &     \text{\small{-0.8}}   & \text{\small{-1.04}}  & \text{\small{-0.95}}  \TBstrut  \\
    \text{\small{Smith09}}     & \text{\small{-3.05} }    &     \text{\small{-2.69}}   & \text{\small{-2,97}}  & \text{\small{-2.90}} \TBstrut  \\
    \text{\small{Smith10}}     & \text{\small{-2.18} }    &     \text{\small{-1.89}}   & \text{\small{-2.14}}   & \text{\small{-2.07}} \TBstrut  \\
     \text{\small{RMSE}}     & \text{\small{ } }    &     \text{\small{0.81}}   & \text{\small{0.53}}  & \text{\small{0.57}} \TBstrut  \\
    \hline \hline
\end{array}
$
\centering
\caption{\label{BE of small Water Clusters}Binding Energies of the 10 Smith dimers using Q-AMOEBA compared to the AMOEBA03 model and ab initio references. The table distinguishes between Q-AMOEBA (PIMD) and Q-AMOEBA (adQTB) depending on which method was used to include NQEs in the calibration of the model. CCSD(T) results come from reference \cite{https://doi.org/10.1002/(SICI)1096-987X(19980730)19:10<1179::AID-JCC6>3.0.CO;2-J}.}
\end{table*}
Over all the configurations explored, both Q-AMOEBA models give very similar and accurate results.
In particular, Q-AMOEBA consistently yields more accurate BE for the Smith dimers than AMOEBA03 with Root Mean Squared Errors (RMSE) in kcal/mol of 0.53 and 0.57 compared to 0.81 for AMOEBA03.
For water clusters of intermediate sizes ($3\leq n \leq 8$), Q-AMOEBA slightly overestimates BEs while remaining within a 0.5 kcal/mol per monomer range. 
The accuracy of Q-AMOEBA improves for the largest clusters, yielding an overall smaller RMSE than AMOEBA03 (overall improvement of approximately 0.9 and 0.2 kcal.mol$^{-1}$). This agreement should translate into a better representation of condensed phase properties.
 \begin{table*}
    \centering
    \resizebox{1\textwidth}{!}{
    $
    \begin{array}{*{12}{c}}
    \hline \hline     
    \text{\small{(H$_2$O)$_n$}}     & \text{\small{geometry} }    &     \text{\small{QM}}   & \text{\small{AMOEBA03}}    & \text{\small{Q-AMOEBA (PIMD)}}  & \text{\small{Q-AMOEBA (adQTB)}}\TBstrut  \\
    \hline
    \text{n=3} & \text{\small{cyclic}} &  \text{\small{-15.74}} &  \text{\small{-15.03}} &       \text{\small{-16.05}}  & \text{\small{-16.11}}  \TBstrut \\
    \text{n=4} & \text{\small{cyclic}} &  \text{\small{-27.40}} &  \text{\small{-27.63}} &       \text{\small{-28.98}}  &       \text{\small{-29.32}}\TBstrut \\
    \text{n=5} & \text{\small{cyclic}} &  \text{\small{-35.93}} &  \text{\small{-36.38}} &     \text{\small{-38.19}}  &       \text{\small{-38.72}}\TBstrut \\
    \text{n=6} & \text{\small{prism}} &  \text{\small{-45.92}} &  \text{\small{-45.71}} &       \text{\small{-47.88}}  &       \text{\small{-48.36}}\TBstrut \\
    \text{ } & \text{\small{cage}} &  \text{\small{-45.67}} &  \text{\small{-45.82}} &       \text{\small{-47.95}}  &       \text{\small{-48.39}} \TBstrut \\
    \text{ } & \text{\small{bag}} &  \text{\small{-44.30}} &  \text{\small{-44.79}} &       \text{\small{-46.97}}  &       \text{\small{-47.53}}\TBstrut \\
    \text{ } & \text{\small{cyclic chair}} &  \text{\small{-44.12}} &  \text{\small{-44.62}} &  \text{\small{-46.88}}  &       \text{\small{-47.60}}\TBstrut \\
    \text{ } & \text{\small{book1}} &  \text{\small{-45.20}} &  \text{\small{-45.6}} &       \text{\small{-47.8}}  &       \text{\small{-48.39}}\TBstrut \\
    \text{ } & \text{\small{book2}} &  \text{\small{-44.90}} &  \text{\small{-45.37}} &       \text{\small{-47.55}}  &       \text{\small{-48.11}}\TBstrut \\
    \text{ } & \text{\small{cyclic boat1}} &  \text{\small{-43.13}} &  \text{\small{-43.78}} &  \text{\small{-45.79}}  &       \text{\small{-46.65}}\TBstrut \\
    \text{ } & \text{\small{cyclic boat2}} &  \text{\small{-43.07}} &  \text{\small{-43.84}} &  \text{\small{-46.04}}  &       \text{\small{-46.73}} \TBstrut \\
    \hline
    \text{n=8 } & \text{\small{S4}} &  \text{\small{-72.70}} &  \text{\small{-72.34}} &       \text{\small{-75.71}}  &       \text{\small{-76.46}} \TBstrut \\
    \text{ } & \text{\small{D2d}} &  \text{\small{-72.70}} &  \text{\small{-72.39}} &       \text{\small{-75.75}}  &       \text{\small{-76.50}}\TBstrut \\
    \text{n=11} & \text{\small{434}} &  \text{\small{-105.72}} &  \text{\small{-101.74}} &       \text{\small{-106.40}}  &       \text{\small{-107.52}}\TBstrut \\
    \text{ } & \text{\small{515}} &  \text{\small{-105.18}} &  \text{\small{-102.08}} &       \text{\small{-106.85}} &       \text{\small{-108.11}} \TBstrut \\
    \text{ } & \text{\small{551}} &  \text{\small{-104.92}} &  \text{\small{-101.85}} &       \text{\small{-106.61}} &       \text{\small{-107.86}} \TBstrut \\
    \text{ } & \text{\small{443}} &  \text{\small{-104.76}} &  \text{\small{-101.86}} &       \text{\small{-106.63}}  &       \text{\small{-107.78}}\TBstrut \\
    \text{ } & \text{\small{4412}} &  \text{\small{-103.97}} &  \text{\small{-101.41}} &       \text{\small{-106.26}}  &       \text{\small{-107.50}}\TBstrut \\
    \text{n=16} & \text{\small{boat-a}} &  \text{\small{-170.80}} &  \text{\small{-161.21}} &       \text{\small{-168.83}}  &       \text{\small{-170.57}}\TBstrut \\
    \text{ } & \text{\small{boat-b}} &  \text{\small{-170.63}} &  \text{\small{-161.54}} &       \text{\small{-169.12}}  &       \text{\small{-170.92}} \TBstrut \\
    \text{ } & \text{\small{antiboat}} &  \text{\small{-170.54}} &  \text{\small{-162.26}} &       \text{\small{-169.98}}  &       \text{\small{-171.95}}\TBstrut \\
    \text{ } & \text{\small{ABAB}} &  \text{\small{-171.05}} &  \text{\small{-161.18}} &       \text{\small{-168.69}}  &       \text{\small{-170.32}}\TBstrut \\
    \text{ } & \text{\small{AABB}} &  \text{\small{-170.51}} &  \text{\small{-160.98}} &       \text{\small{-168.51}}  &       \text{\small{-170.13}}\TBstrut \\
    \text{n=17} & \text{\small{sphere}} &  \text{\small{-182.54}} &  \text{\small{-173.02}} &       \text{\small{-181.21}}  &       \text{\small{-183.32}} \TBstrut \\
    \text{ } & \text{\small{5525}} &  \text{\small{-181.83}} &  \text{\small{-172.32}} &       \text{\small{-180.42}}  &       \text{\small{-182.46}} \TBstrut \\
    \text{n=20} & \text{\small{dodecahedron}} &  \text{\small{-200.10}} &  \text{\small{-197.03}} &       \text{\small{-206.1}} &       \text{\small{-208.54}} \TBstrut \\
    \text{ } & \text{\small{fused cubes}} &  \text{\small{-212.10}} &  \text{\small{-205.76}} &       \text{\small{-215.35}}  & \text{\small{-217.42}} \TBstrut \\
    \text{ } & \text{\small{face sharing prisms}} &  \text{\small{-215.20}} &  \text{\small{-206.13}} &       \text{\small{-215.74}}   & \text{\small{-218.05}} \TBstrut \\
    \text{ } & \text{\small{edge sharing prisms}} &  \text{\small{-218.10}} &  \text{\small{-208.46}} &       \text{\small{-218.40}}   & \text{\small{-220.85}}\TBstrut \\
    \text{RMSE} & \text{\small{ }} &  \text{\small{ }} &  \text{\small{3.33}} &       \text{\small{2.36}}   & \text{\small{3.12}} \TBstrut \\
    \hline \hline
\end{array}
$
}
\caption{\label{BE of big Water Clusters}Binding Energies of trimer to 20-mer of water clusters computed using Q-AMOEBA compared to AMOEBA03 and ab initio. The table distinguishes between Q-AMOEBA (PIMD) and Q-AMOEBA (adQTB) depending on which method was used to include NQEs in the calibration of the model. QM reference correspond to CCSD(T) for $n \leq 6$ \cite{Bates2009-om,Ren2003} and to MP2 for $n \geq 8$ \cite{doi:10.1063/1.1626624,Yoo2010,doi:10.1063/1.1767519}. }
\end{table*}
\subsection{Structural properties of water}
\begin{figure}
    \centering
    \includegraphics[width=\textwidth, height=0.65\textheight,keepaspectratio]{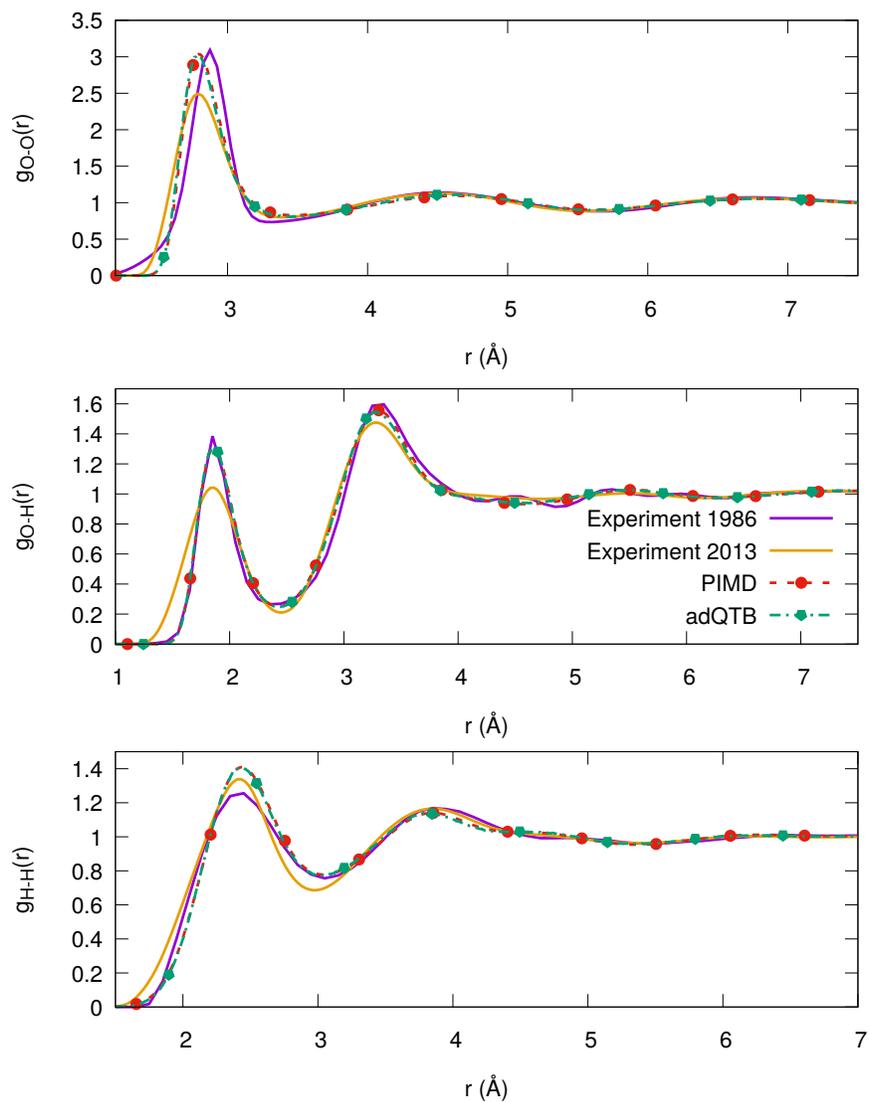}
    \centering
    \caption{\label{RdFs} RDFs computed at 298.15K with Q-AMOEBA compared to the X-ray (2013) \cite{doi:10.1063/1.4790861} and Neutron (1986) \cite{SOPER2000121} diffraction experiments.  }
\end{figure}

Radial Distribution Function (RDFs) are a good indicator of the local molecular structure in the liquid phase both in terms of the different peak positions and their width. They were computed from NVT simulations at 298.15 K and the corresponding equilibrium volume. 
In Figure \ref{RdFs}, the PIMD (dash red curve) and adQTB (dot and dash green curve) radial distributions are compared to the experimental data obtained from X-ray scattering (2013) \cite{doi:10.1063/1.4790861} and neutron diffraction (1986) \cite{SOPER2000121}. These data were not included in our parametrization and hence were used as a first test.

Both Q-AMOEBA results are almost indistinguishable and overall in close agreement with experiments. The difference between simulation and experimental curves are of the same order as the discrepancies within experimental data. 
For instance, the first peak of the Oxygen-Oxygen RDF is located at 2.73~\AA\  in the X-ray scattering experiment whereas it is at  2.80~\AA \text{ } in the neutron diffraction one. It is well reproduced by the two Q-AMOEBA models, with peaks located respectively at   2.80~\AA \text{ } (PIMD) and 2.79~\AA \ (adQTB).

Intramolecular structure is illustrated by the average O-H distance and H-O-H angle as shown in table \ref{Static properties}. NQEs tend to slightly increase the average O-H distance and H-O-H angle of water molecules in the liquid phase.
This is why both sets of parameters have a reduced ideal angle compared to the initial set of parameters (106.2$^{\circ}$ in adQTB and 107.2$^{\circ}$ in PIMD compared to 108.2$^\circ$ in AMOEBA03). However, the PIMD values of table \ref{Static properties} are a little overestimated whereas this effect was compensated in our adQTB model by decreasing further the associated equilibrium parameters (from 0.9572 to 0.9472~\AA~ for the O-H distance and 108.2$^{\circ}$ to 106.2$^{\circ}$ for the H-O-H angle). Interestingly, the modeling of the average value of the H-O-H angle in liquid phase compared to its average value in gas phase is known to be problematic with classical force fields and requires to artificially increase the equilibrium parameter \cite{Ren2003}. It has been shown that this can be traced back to the absence of charge-flux effect \cite{AMOEBA+2}. The former effect of NQEs on these averages highlights the need to explicitly include them to design accurate models of water.
 \begin{table*}
    \centering
    $
    \begin{array}{*{12}{c}}
    \hline \hline     
    \text{\small{ }}        &     \text{\small{PIMD (32 beads)}}   &   \text{\small{PIMD (64 beads)}} & \text{\small{adQTB}} & \text{\small{Expt.}}\TBstrut  \\
    \hline
   \text{\small{r$_{OH}$(\AA)}}     &     \text{\small{0.985}}     &     \text{\small{0.985}}    & \text{\small{0.975}} & \text{\small{0.97}} \TBstrut  \\
     \text{\small{$\theta_{HOH}$(deg)}}      &  \text{\small{106.08}}   &     \text{\small{106.10}}  & \text{\small{104.95}} & \text{\small{105.1}} \TBstrut  \\
    \hline \hline
\end{array}
$
\caption{\label{Static properties}Average O-H bond length and H-O-H angle while using the Q-AMOEBA parameters compared to the experimental results \cite{SOPER2000121}. }
\end{table*}



\subsection{Infrared spectra}
\begin{figure*}[!]
    \centering
    \includegraphics[scale=0.1]{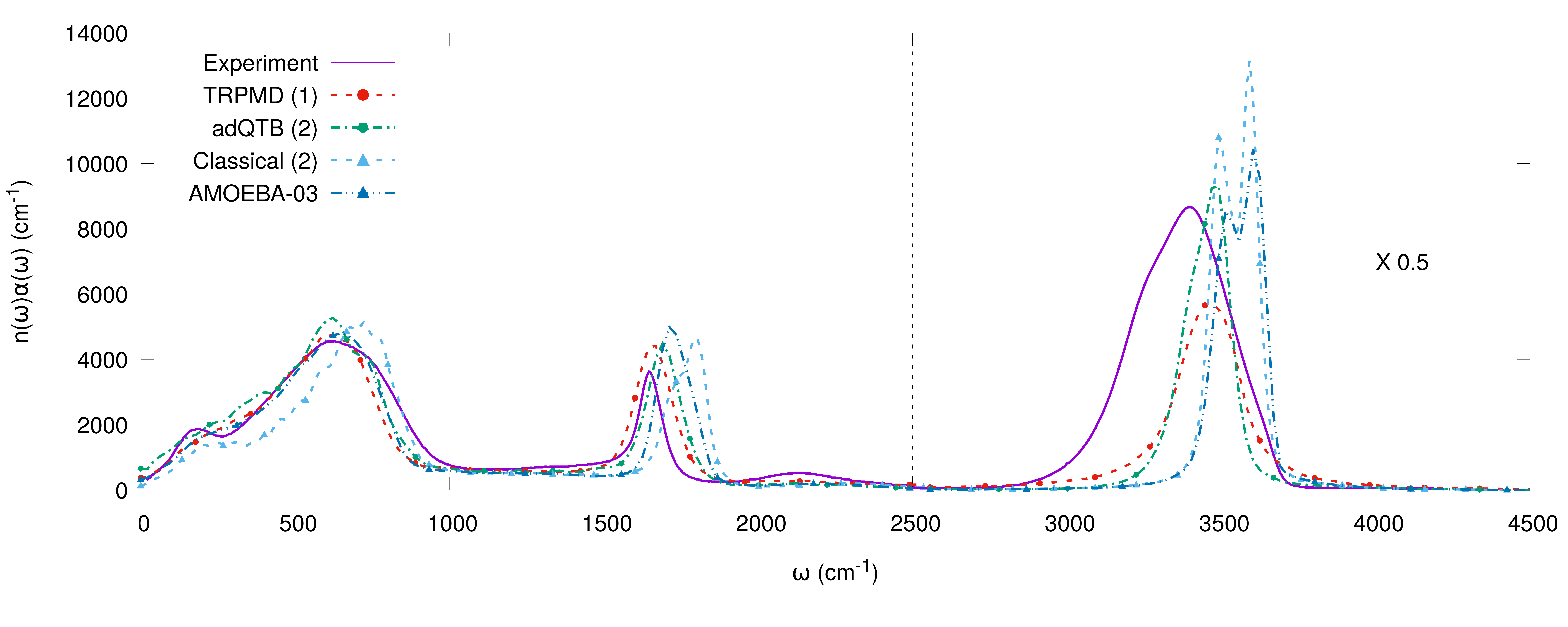}
    \centering
    \caption{\label{IR figure} IR absorption spectra computed at 300~K and $\rho=0.997$ g.cm$^{-3}$ with the Q-AMOEBA water model the adQTB and PIMD methods with their respective set of parameters (green and red curves, respectively). The \textit{classical} curve (light blue) is obtained from classical MD simulations with the Q-AMOEBA (adQTB) model. They are compared with AMOEBA03 (classical MD) and experimental data \cite{Bertie1989}. The right part of the plot, corresponding to the stretching mode is multiplied by 0.5. The number in parenthesis refers to the set of force field parameters used in the simulations: (1) PIMD; (2) adQTB (see text).}
\end{figure*}
The infrared absorption spectrum can be related to the total dipole derivative autocorrelation function as \cite{zwanzig2001nonequilibrium,benson2020quantum}:
\begin{equation}
    n(\omega)\alpha(\omega) = \frac{\pi\beta}{3V\epsilon_0}C_{\dot{\mu}\dot{\mu}}(\omega)
\end{equation}
with $n(\omega)$ the refractive index, $\alpha(\omega)$ the Beer-Lambert absorption coefficient, $\beta=1/k_BT$ the inverse thermal energy, $V$ the system's volume and $C_{\dot{\mu}\dot{\mu}}(\omega)$ the Kubo autocorrelation spectrum of the total dipole time derivative $\dot{\mu}$. In the AMOEBA framework, the total dipole moment is estimated as:
\begin{equation}\label{eq:total_dipole}
    \vb{\mu} = \sum_{i=1}^{N_\ttiny{at}}~\big(q_i\vb{r}_i + \vb{\mu}_i^0 + \vb{\mu}_i^\ttiny{ind}\big)
\end{equation}
where $q_i$ are permanent charges located on the atom's position $\vb{r}_i$, $\vb{\mu}_i^0$ are permanent dipoles (which magnitudes are fixed but rotate with the water molecules) and $\vb{\mu}_i^\ttiny{ind}$ are induced dipoles that are obtained 
at each step of the dynamics via a minimization procedure \cite{C7SC04531J}. Since no analytical form for the time derivative of the induced dipole is available, it is estimated using finite differences of the total dipole moment over the trajectory.

The IR spectrum is directly evaluated in path integrals simulations in the framework of TRPMD \cite{rossi2014remove} while it is recovered from adQTB simulations through the following relation in the Fourier domain \cite{ple2021anharmonic}:
\begin{equation}
    C_{\dot{\mu}\dot{\mu}}(\omega) \approx \frac{\tanh(\beta\hbar\omega/2)}{\beta\hbar\omega/2}C^\ttiny{adQTB}_{\dot{\mu}\dot{\mu}}(\omega)
\end{equation}
In practice, it is directly estimated from the Fourier transform of the dipole derivative trajectory (averaged over the beads in TRPMD simulations) according to the Wiener-Khinchin theorem. Furthermore, in the adQTB method, the relatively high friction coefficient (that is required to compensate ZPEL) tends to broaden spectral lineshapes so that it is necessary to use the deconvolution procedure of ref.~\cite{rossi2018fine} to improve the spectrum.
\par
Figure~\ref{IR figure} shows the IR spectra calculated using TRPMD and adQTB compared to the experimental spectrum (extracted from ref.~\cite{doi:10.1063/1.3167790}), classical AMOEBA03 spectrum and the classical spectrum obtained from a simulation using the parameters optimized for adQTB. 
The spectra from both quantum simulations are very similar in their peak positions and are in good agreement with the experimental data. As already well documented \cite{Ceriotti2016,benson2020quantum,doi:10.1021/acs.jpclett.1c01722}, including NQEs shifts the intramolecular peaks (bending around $1600 \,\text{cm}^{-1}$ and stretching around $3500 \,\text{cm}^{-1}$) towards lower frequencies. 
Relative intensities between the spectral features are similar in quantum and classical simulations and agree well with experimental results (although, we can note that the TRPMD spectrum for the stretching peak is broadened compared to the other simulations, which is a well known discrepancy of this method). 
The low-frequency features are very similar for AMOEBA03, TRPMD and adQTB (although slightly more intense for the adQTB) and are red-shifted compared to the results obtained in classical simulations with Q-AMOEBA (light blue), showing the impact of NQEs even at those low frequencies. The general shape of these low-frequency features are in good agreement with the experimental spectrum and even display some sub-structure at around $200 \,\text{cm}^{-1}$ that is due to slow induced-dipole dynamics \cite{doi:10.1063/1.3167790,impey1982spectroscopic} and is absent for non-polarizable models, even when including NQEs \cite{doi:10.1063/1.3167790}.

\subsection{Thermodynamic properties}
\label{sec:thermo_prop}
\begin{figure}
    \centering
    \includegraphics[scale=0.185]{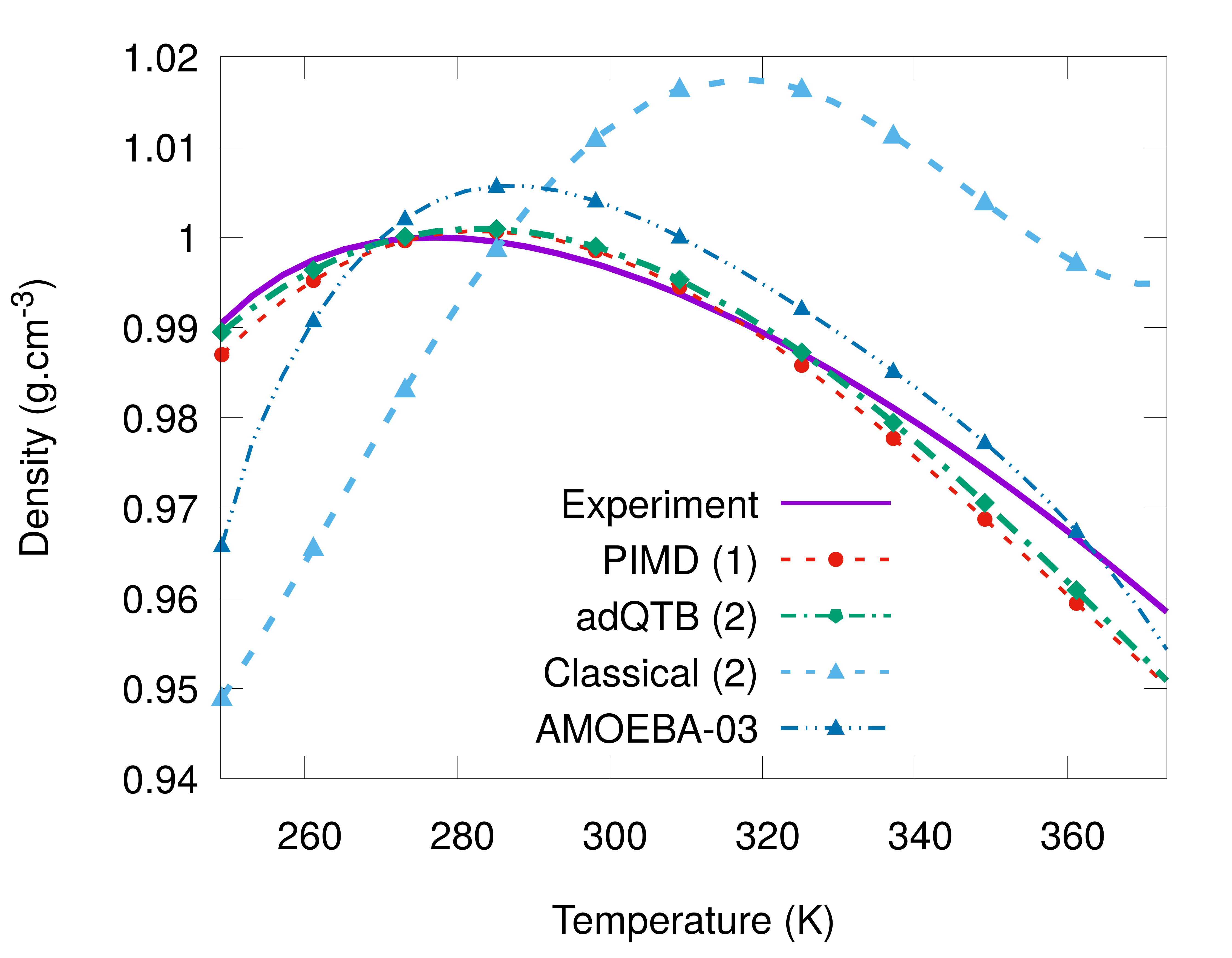}
    \centering
    \caption{\label{Density} Density of liquid water at $P=1$~atm as a function of the temperature for different AMOEBA models. The dashed line represents the density while using the Q-AMOEBA PIMD method whereas the dot-dashed lines are obtained with the Q-AMOEBA adQTB method. Experimental data come from \cite{Kell1975}. The number in parenthesis refers to the set of force field parameters used in the simulations: (1) PIMD; (2) adQTB (see text). 
    }
\end{figure}

Thermodynamic properties of the two Q-AMOEBA set of parameters were computed from temperatures ranging from 249.15K to 369.15 K under 1 atm. As shown in Figure \ref{Density}, the characteristic bell shape of the density versus temperature curve is well captured by the Q-AMOEBA force fields. The deviations from the experimental result are minor, with average densities at ambient temperature of 0.998 and 0.999 g.cm$^{-3}$ for the PIMD and adQTB methods respectively, compared to the experimental value of 0.997 g.cm$^{-3}$ (less than 0.2$\%$ difference). In both high and low ranges of temperature, the density is slightly underestimated by a maximum of  0.85$\%$ at 369.15K. 
The Figure also shows the curve obtained from a classical MD simulation using the Q-AMOEBA (adQTB) set of parameters (light blue curve). It differs significantly from the AMOEBA03 classical reference, which shows that the changes made to the force field parameters have a sizeable impact on the density. Even more importantly, the classical MD curve displays strong discrepancies with the corresponding adQTB and PIMD results: the temperature of maximum density $T_{MD}$ is shifted to higher values while the density is significantly reduced at low temperature and increased at high temperatures with respect to the results obtained including NQEs. To explain this impact of NQEs, we come back to the peculiarities of liquid water that explain the unusual bell shape of the density curve. The latter is related to the tetrahedral arrangement of the water molecules, which tends to produce a loosely packed local structure. For $T<T_{MD}$, increasing the temperature allows hydrogen bonds to strain and break more easily, which in turns enables larger deviations from the tetrahedral order with a better packing efficiency and therefore an increase of the density. On the other hand, for $T>T_{MD}$, a competitive process tends to dominate: the breaking of H-bonds tends to increase nearest-neighbors distances which leads to the more usual thermal expansion observed at high temperature~\cite{stillinger1980water_revisited}. Within the Q-AMOEBA model, we observe that the inclusion of NQEs tends to increase the density for $T<T_{MD}$ and to decrease it above $T_{MD}$, which indicates that NQEs lead to an overall weakening of the hydrogen bonds over the whole temperature range. This is consistent with experimental observations of isotope effects in liquid water, most of which suggest that hydrogen bonds are weaker in H$_2$O than in D$_2$O~\cite{Ceriotti2016}, the latter isotope being heavier hence \textit{more classical}. In particular, the experimental $T_{MD}$ shifts from 277.15~K for H$_2$O to 284.32~K for D$_2$O and even 286.55~K for T$_2$O, which agrees qualitatively with the observed shift between the quantum and classical MD curves within the Q-AMOEBA model. 
 \begin{table*}
    \centering
    \resizebox{1\textwidth}{!}{
    $
    \begin{array}{*{12}{c}}
    \hline \hline     
    \text{\small{ }}     & \text{\small{Bond stretching} }    &     \text{\small{Angle bending}}   & \text{\small{Urey-Bradley}} & \text{\small{Van der Waals}} & \text{\small{Atomic Multipoles}} & \text{\small{Polarization}}  & \text{\small{Non-Bonded}} \TBstrut  \\
    \hline
    \text{\small{Classical (2)}}  & \text{\small{0.75} }  & \text{\small{0.43} }  & \text{\small{-0.03} }  & \text{\small{5.48} }  & \text{\small{-11.43} }  & \text{\small{-5.19} }  & \text{\small{-11.14} }\TBstrut  \\
    \text{\small{adQTB (2)}}     & \text{\small{5.55} }    &     \text{\small{1.22}}   & \text{\small{-0.11}} & \text{\small{4.37}} & \text{\small{-9.95}} & \text{\small{-4.43}} & \text{\small{-10.01}} \TBstrut  \\
    \text{\small{PIMD (1) (32 beads)}}     & \text{\small{5.43}}    &     \text{\small{1.19}}   & \text{\small{-0.12}} & \text{\small{4.32}} & \text{\small{-9.87}} & \text{\small{-4.35}} &\text{\small{-9.9}}\TBstrut  \\
     \text{\small{PIMD (1) (64 beads)}}     & \text{\small{5.57}}    &     \text{\small{1.20}}   & \text{\small{-0.12}} & \text{\small{4.33}} & \text{\small{-9.88}} & \text{\small{-4.38}} & \text{\small{-9.93}}\TBstrut  \\
    \hline \hline
\end{array}
$
}
\caption{\label{Observables_values} Average energy contributions of Q-AMOEBA obtained with the different methods at 298.15~K. The classical row corresponds to the adQTB set of parameter. All energies are given in kcal.mol$^{-1}$ per water molecule and obtained from 1~ns simulations at constant volume corresponding to the density $\rho=0.997$ g.cm$^{-3}$. The number in parenthesis refers to the set of force field parameters used in the simulations: (1) PIMD; (2) adQTB (see text).}
\end{table*}

This reasoning is further confirmed by the study of the different non-bonded energy contributions of the Q-AMOEBA force field in Table \ref{Observables_values}. PIMD values are superior by approximately 1.0 kcal/mol per water molecule at 298.15K to their classical counterpart, a trend which is consistent over the whole temperature range.
Notably, the impact of NQEs on the hydrogen bond strength strongly depends on the underlying water model\cite{li2022static}. For example in the q-TIP4P/f model, densities computed in PIMD or in classical MD are very similar \cite{doi:10.1021/acs.jpclett.1c01722}, whereas in the TTM2.1-F FF both curves are overestimated \cite{doi:10.1063/1.2759484} and the PIMD density is lower than its classical counterpart at all temperatures.

\begin{figure}
    \centering
    \includegraphics[scale=0.2]{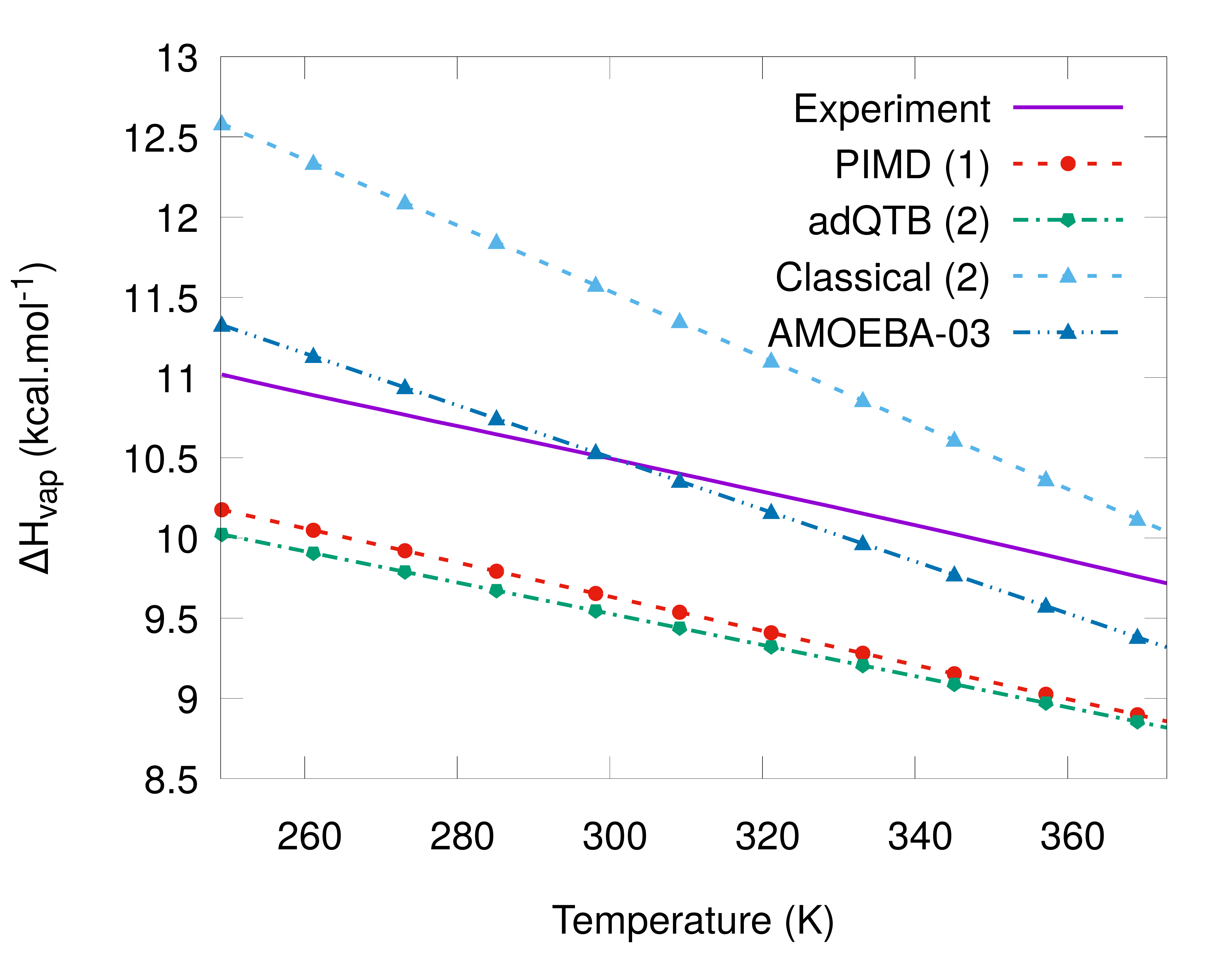}
    \centering
    \caption{\label{DeltaHvap} Enthalpy of vaporization at $P=1$~atm as a function of the temperature with the different models. The dashed line represents the PIMD results and the dot-dashed lines the adQTB ones. The number in parenthesis refers to the set of force field parameters used in the simulations: (1) PIMD; (2) adQTB (see text). Experimental data (continuous line) are taken from \cite{doi:10.1063/1.1461829}.}
\end{figure}

It has been recognized that NQEs also play an important role in determining the enthalpy of vaporization $\Delta H_{vap}$ of water and that they should be taken into account explicitly for accurate prediction of this observable\cite{doi:10.1063/1.4967719,doi:10.1063/1.1356002}. 
Assuming water vapor to be an ideal gas, one may find:
\begin{equation}
    \Delta H_{vap}=E_g-E_{liq}+P(V_g-V_l)
\end{equation}
where $E_g$ is the average total energy in the gas phase and $E_{liq}$ the average total energy per molecule in the liquid phase. As shown in Figure \ref{DeltaHvap}, classical enthalpy of vaporization is overestimated and decreases with the inclusion of NQEs.
The Q-AMOEBA classical result (dot light blue) yields $\Delta H_{vap}=11.5$ kcal.mol$^{-1}$ at 298.15~K which is $\sim$0.5  kcal.mol$^{-1}$ higher than the reference. On the other hand the PIMD and adQTB results underestimate the $\Delta H_{vap}$ by approximately $\sim$0.8 kcal.mol$^{-1}$, similarly to other sophisticated models~\cite{medders2014development}, while capturing the correct slope of the curve with increasing temperature. Since the correct slope is recovered, some hypothesis tracking back the small resulting Q-AMOEBA discrepancies (compared to experiment) to the reference ab initio computations can be introduced. First, as we discussed the impact of NQEs on the hydrogen bond strength, it is highly probable that our CCSD(T) reference level does not fully reflect the experimental high complexity of the potential energy surfaces of water dimers and clusters. Indeed, at this level of accuracy, small changes in gas phase reference computations could still impact the predicted condensed phase properties. If the explicit inclusion of triple excitations clearly increases the strength of interactions (\cite{Benchdimereccsd}), it has been shown that CCSDTQ approaches yield to relatively small but non-negligible improvement in the optimized CCSD(T) geometry and interaction energy for the water dimer~\cite{CCSDTQ}. Furthermore, one can guess that the simple Q-AMOEBA functional form probably reach its limit in capturing highly complex electron density delocalization/overlap effects that would require the inclusion of higher-order excitations. Models like SIBFA~\cite{Naseem-Khan2022} or AMOEBA+ \cite{AMOEBA+,AMOEBA+2} that include additional overlap/delocalization effetcs such as electrostatic penetration, charge transfer, etc.., could be potentially better able to capture such non-trivial interactions. 

The validity of Q-AMOEBA was further studied by computing other equilibrium properties of liquid water which were not included in the objective function of FB: the isobaric heat capacity $c_p$, the thermal expansion coefficient $\alpha_P$, the isothermal compressibility $\kappa_T$ and the dielectric constant $\varepsilon_r$. 
The isobaric heat capacity is defined as:
\begin{equation}
    \label{cp_equation}
    c_p=\left(\frac{\partial H}{\partial T}\right)_P
\end{equation}
where \textit{H} and \textit{T} are the enthalpy and the temperature at a given pressure \textit{P}. In practice, $c_p$ is estimated as the temperature derivative of a fourth-order polynomial interpolating the values of \textit{H} obtained at different temperatures. It is well known that the heat capacity is strongly affected by NQEs. Indeed, at room temperature, the fluctuations of the high-frequency intramolecular modes are almost exclusively due to zero point energy effects and essentially independent of the temperature. Therefore intramolecular energy terms give almost no contribution to the heat capacity, which is not the case in the classical picture, leading to a significant overestimation of $c_p$ when NQEs are not included explicitly \cite{doi:10.1063/1.4967719}. 
\begin{figure}
    \centering
    \includegraphics[scale=0.185]{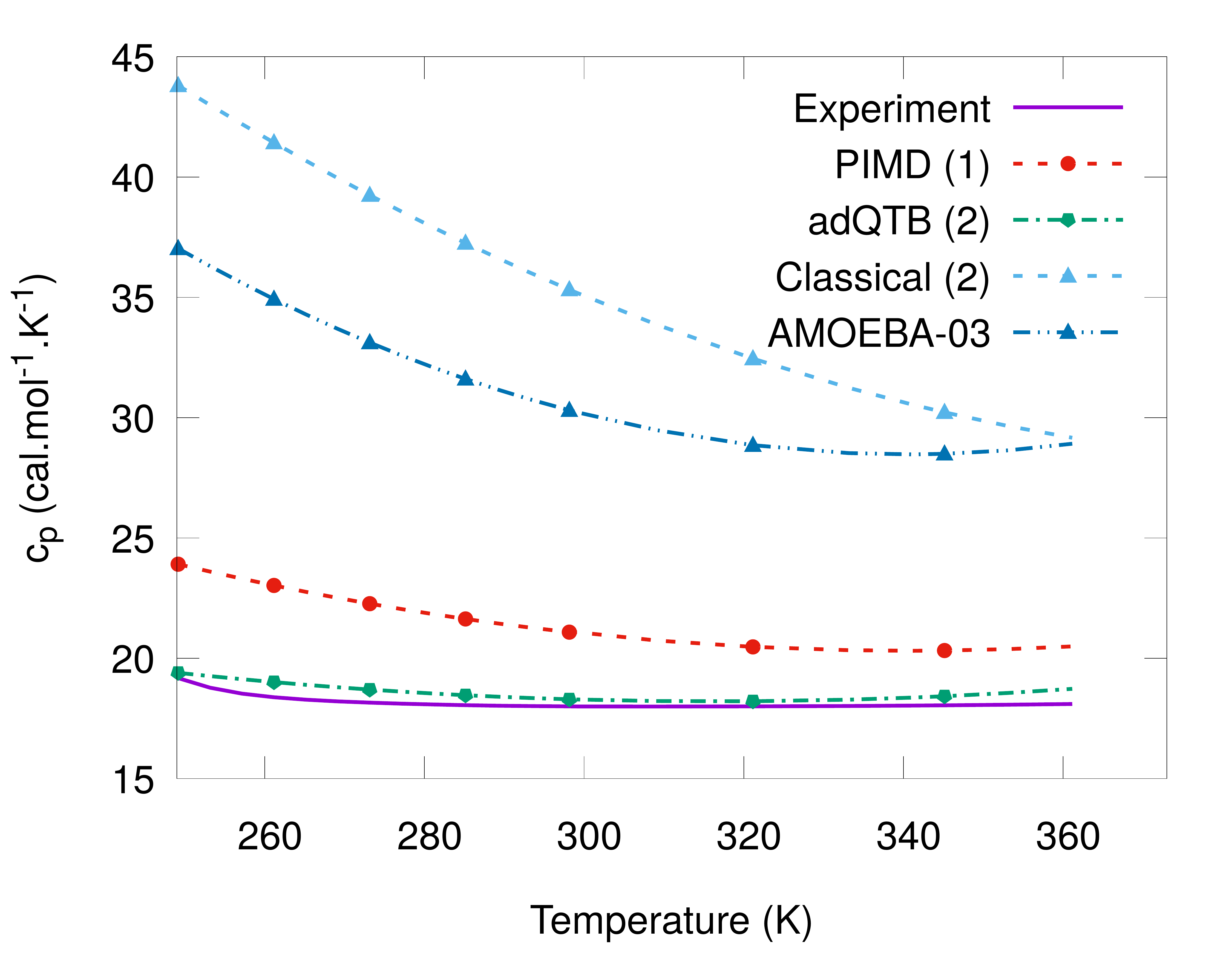}
    \centering
    \caption{\label{cp} Isobaric heat capacity at $P=1$~atm as a function of the temperature with the different models. The dashed line represents the PIMD results and the dot-dashed lines the adQTB ones. The number in parenthesis refers to the set of force field parameters used in the simulations: (1) PIMD; (2) adQTB (see text). Experimental data are taken from \cite{doi:10.1063/1.1461829}. }
\end{figure}
Indeed, as shown in Fig.~\ref{cp}, classical MD simulations overestimate $c_p$ for both Q-AMOEBA and AMOEBA03. On the contrary, including NQEs allows recovering lower values much closer to experiment.

The thermal expansion coefficient $\alpha_P$ is calculated using analytic differentiation of a polynomial fit of the simulated density $\rho(T) $:
\begin{equation}
    \alpha_p=\frac{1}{V}\left (\frac{\partial V}{\partial T}\right )_P=-\frac{\text{d} \ln{ \rho(T)} }{\text{d}T}
\end{equation}
It shows the same trend as described above for the density: classical MD yields too negative results at low temperature, a trend corrected by the inclusion of NQEs. At high temperatures, all the methods behave similarly.
\begin{figure}
    \centering
    \includegraphics[scale=0.185]{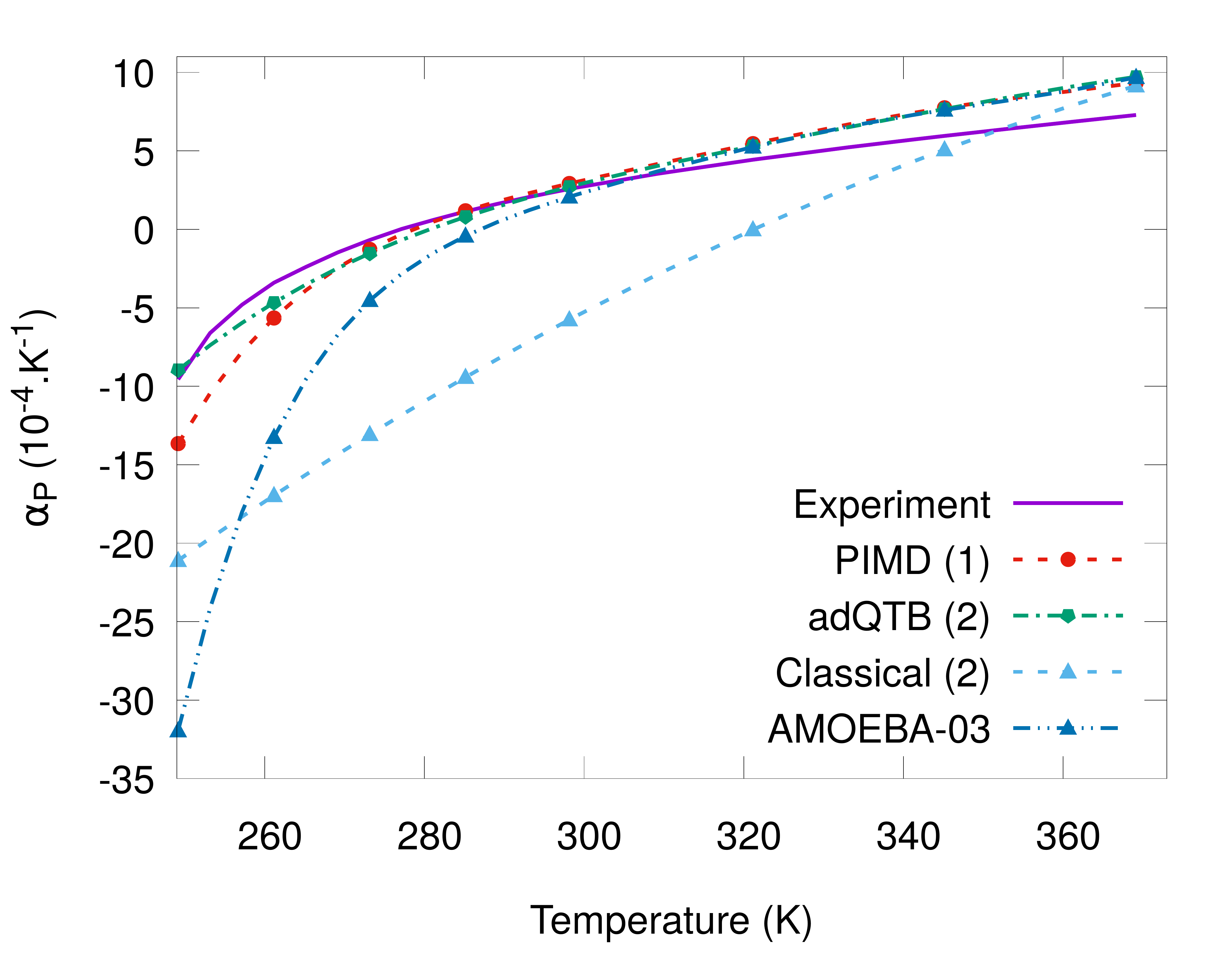}
    \caption{Thermal expansion coefficient at $P=1$~atm as a function of the temperature with the different set of parameters. The number in parenthesis refers to the set of force field parameters used in the simulations: (1) PIMD; (2) adQTB (see text). Experimental data are taken from \cite{Kell1975}. }
    \label{alpha}
\end{figure}

Longer NPT simulations of 10 ns were performed to converge the static dielectric constant and the self-diffusion coefficient and 20 ns for the isothermal compressibility.
The isothermal compressibility, that characterizes the volume change as a response to an applied pressure, can also be related to the volume fluctuations in an NPT simulation:
\begin{equation}
    \kappa_T=-\frac{1}{V}\left(\frac{\partial\,V}{\partial\,P}\right)_{T,N}=\frac{1}{k_B\,T}\frac{\langle V^2 \rangle\,-\,\langle V \rangle^2}{\langle V \rangle}
\end{equation}
\begin{figure}
    \centering
    \includegraphics[scale=0.185]{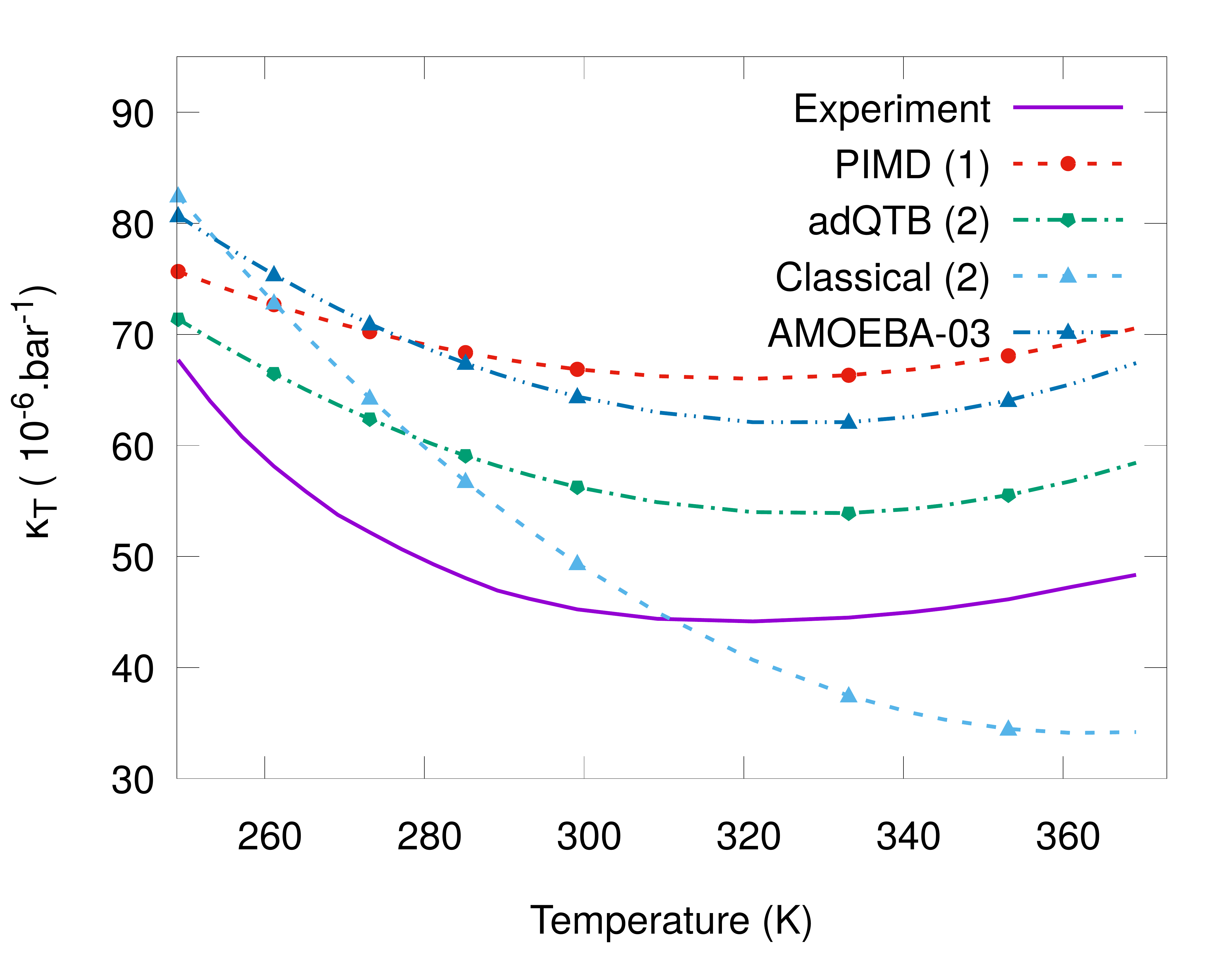}
    \caption{Isothermal compressibility at $P=1$~atm as a function of the temperature with the different models. The number in parenthesis refers to the set of force field parameters used in the simulations: (1) PIMD; (2) adQTB (see text). Experimental data are taken from \cite{kell1975density}.}
    \label{kt}
\end{figure}
Simulations with classical nuclei in the Q-AMOEBA  model tend to amplify the variations of this observable with $T$, and to underestimate its value at high temperature. The explicit inclusion of NQEs reduces the variations of $\kappa_T$ over the explored temperature range, and yield a shape that is very similar to the experimental curve, though with slightly larger values (particularly for the PIMD model). Overall, the results of the adQTB model are significantly improved with respect to that of the original AMOEBA03, whereas the PIMD ones are similar to it, which highlights that the original FF implicitly takes into account NQEs in its parametrization. 
\par
The static dielectric constant is calculated from the fluctuations of the total dipole moment as:
\begin{equation}
    \varepsilon_r=1+\frac{4\,\pi}{3\,k_B\,T \langle V \rangle}\left ( \langle \mu^2 \rangle - \langle \mu \rangle \cdot \langle \mu \rangle \right)
\end{equation}
with $\langle \mu \rangle$  the average total dipole moment defined in eq.~\eqref{eq:total_dipole} and $\langle V \rangle$ the average volume of the simulation box. 
\begin{figure}
    \centering
    \includegraphics[scale=0.185]{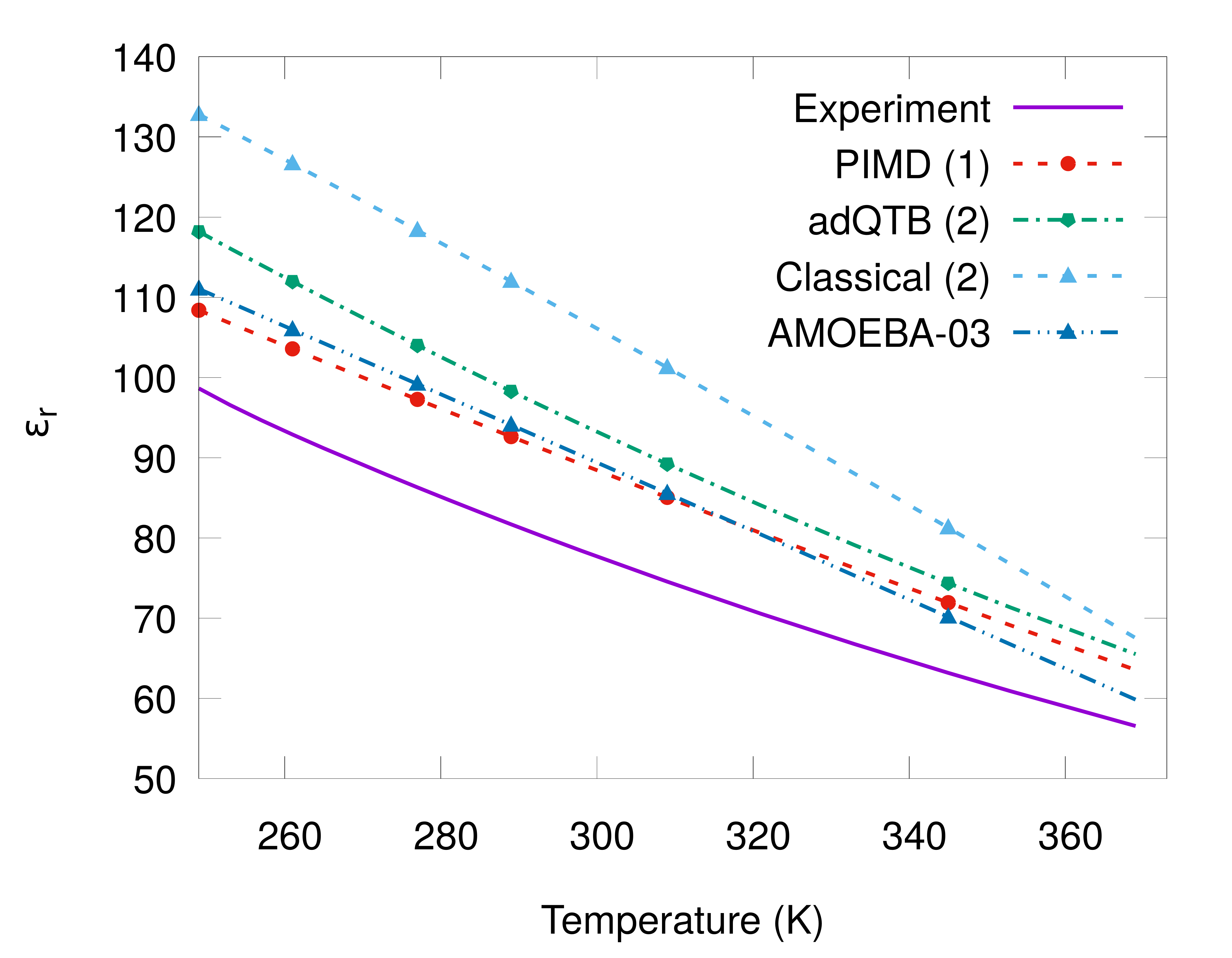}
    \caption{Static dielectric constant at $P=1$~atm as a function of the temperature with the different set of parameters. The number in parenthesis refers to the set of force field parameters used in the simulations: (1) PIMD; (2) adQTB (see text). Experimental data are taken from \cite{doi:10.1063/1.1461829}.}
    \label{diel}
\end{figure}
The results are satisfactory for $\varepsilon_r$ with Q-AMOEBA and show the impact of NQEs on this property: they significantly reduce $\varepsilon_r$ at low temperature and have a smaller influence when T increases. 
\par
The self-diffusion coefficient was evaluated at 298.15K under 1~atm using the Einstein equation:
\begin{equation}
    D_0= \lim\limits_{t \to \infty}\frac{ \text{d}}{\text{d}\,t}\langle \left | r(t)\,-\,r(t_0)|^2\right \rangle
\end{equation}
The mild Langevin thermostat (friction of $\gamma=1\,\text{ps}^{-1}$) applied to the centroid in the TRPMD method has been shown to have only a small effect on the computed diffusion  whereas the adQTB method requires larger frictions ($\gamma=20\,\text{ps}^{-1}$) which affects the diffusion \cite{Maginn_Messerly_Carlson_Roe_Elliot_2018,doi:10.1021/acs.jpclett.1c01722}.
The PIMD (TRPMD) value is 2.29 $\pm 0.01   \times$ $10^5$ cm$^2$.s$^{-1}$ which is in excellent agreement with the experimental value of 2.30 $\times$ $10^5$ cm$^2$.s$^{-1}$ and larger than the associated classical one (0.91 $\pm 0.01   \times$ $10^5$ cm$^2$.s$^{-1}$). The quantum/classical ratio of this quantity has been observed to vary widely among water models \cite{doi:10.1063/1.3167790,doi:10.1063/1.2386157, doi:10.1063/1.2759484}. On the other hand, the adQTB value is 0.78 $\pm 0.01 \times$ $10^5$ cm$^2$.s$^{-1}$ which is around three times lower than the expected value due to the large $\gamma$ of the Langevin thermostat\cite{doi:10.1021/acs.jpclett.1c01722}. The slower diffusion can be associated with a less efficient sampling \cite{berendsen1981interaction}, but this is largely compensated by the considerable speedup of the adQTB method compared to path integral ones.
\begin{figure}
    \centering
    \includegraphics[scale=0.185]{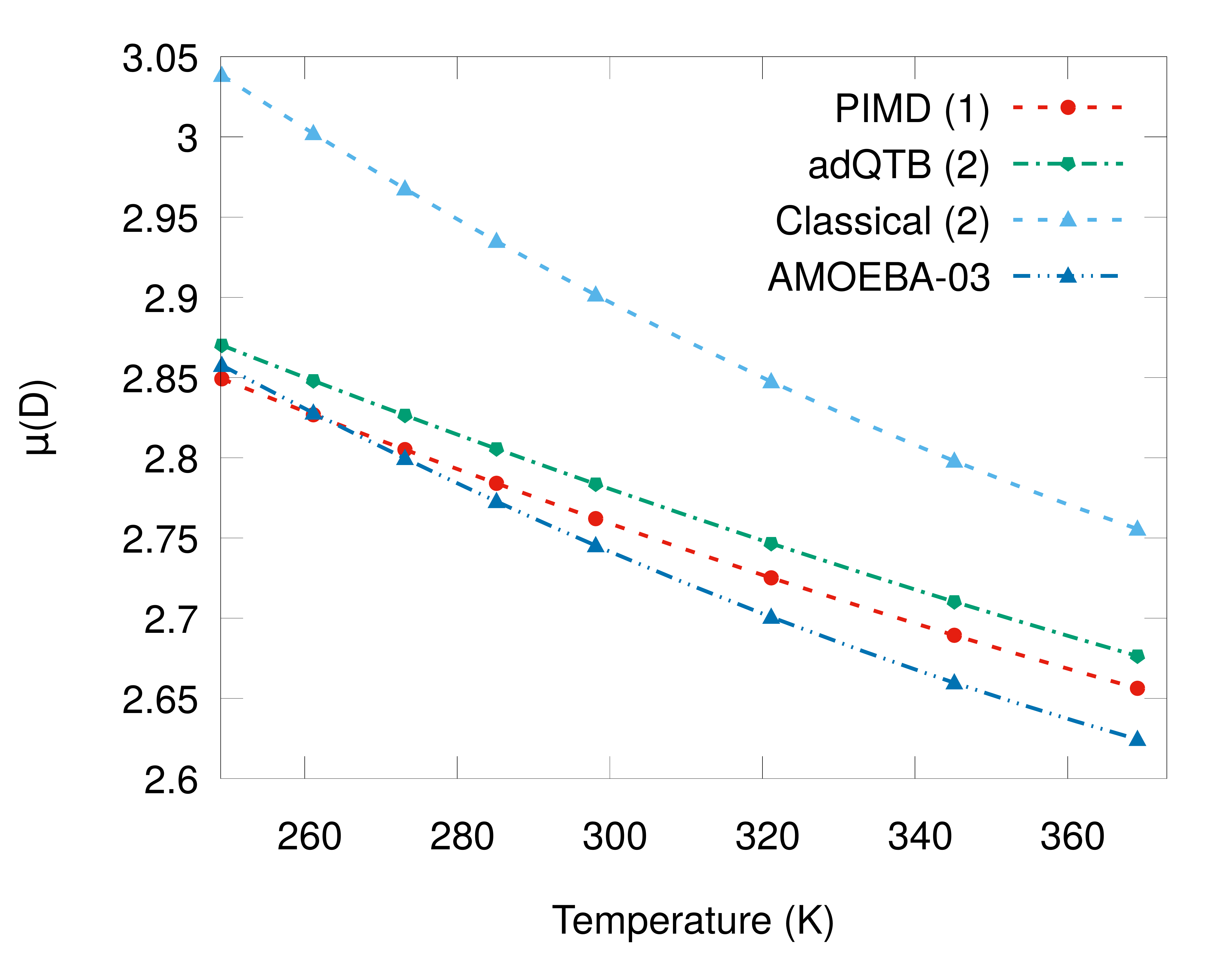}
    \centering
    \caption{\label{dipole_moment} Molecular dipole moments at $P=1$ atm as a function of the temperature with the different set of parameters. The number in parenthesis refers to the set of force field parameters used in the simulations: (1) PIMD; (2) adQTB (see text).}
\end{figure}
\par
Finally, the molecular dipole moments were also studied in order to further confirm the previous interpretation regarding the impact of NQEs with respect to the underlying hydrogen bonds strength. Figure \ref{dipole_moment} shows that, in classical MD simulations with the Q-AMOEBA model, the dipole moment is larger than in the original AMOEBA03 force field, indicating stronger hydrogen bonds in the new model. At room temperature, the value obtained from classical simulations with the Q-AMOEBA model is 2.90 D, whereas, explicitly including NQEs via PIMD or adQTB weakens the H-bonds and reduces the dipole moment to 2.76 D or 2.78 D, respectively. These values are close to those obtained in AMOEBA03, and in agreement with \textit{ab initio} simulations and experiments~\cite{badyal2000electron,gubskaya2002total}. 

\subsection{Heavy water}
Isotope effects are an important marker of the relevance of NQEs, since classical thermodynamics predicts different isotopes to have identical thermal equilibrium properties. Therefore, explicit treatment of NQEs is necessary to describe effects such as those arising from the substitution of H by its heavier isotope D\cite{C1CP21863H}. To further validate the Q-AMOEBA approach, we present in Table \ref{D2O_properties} the values for heavy water $\text{D}_2\text{O}$ of some thermodynamic properties previously presented, focusing on those known to be most impacted by the isotopic substitution.
\begin{table}
\begin{tabular}{ccccccc}
\hline \hline
          & \multicolumn{3}{c}{H$_2$ O} & \multicolumn{3}{c}{D$_2$ O} \TBstrut \\ \cline{2-7} 
          & PIMD (1)   & adQTB (2)   & Exp.   & PIMD (1)    & adQTB (2)   & Exp.  \Tstrut \Bstrut \\ \hline
$\rho$    & 0.998       &  0.999       & 0.997  &  1.109       &      1.106  & 1.104  \TBstrut \\
$\Delta$H &  9.65      &   9.55      & 10.52  &      10.25   &     10.28   & 10.85   \\
c$_p$   & 21.087  &    18.297     &   18.002     &   22.724      &    21.885    & 20.148   \Bstrut    \\ 
\hline \hline
\end{tabular}
\caption{\label{D2O_properties} Thermodynamic properties computed at 298.15K under 1 atm pressure for both Q-AMOEBA models for $\text{H}_2\text{O}$ and $\text{D}_2\text{O}$ compared to experimental values \cite{doi:10.1063/1.555561,doi:10.1063/1.1461829}. The density $\rho$ is given in g.cm$^{-3}$, the enthalpy of vaporization $\Delta H_{vap}$ in kcal.mol$^{-1}$ and the isobaric heat capacity $c_p$ in cal.mol$^{-1}$.K$^{-1}$. The number in parenthesis refers to the set of force field parameters used in the simulations: (1) PIMD; (2) adQTB (see text }
\end{table}

Overall, Q-AMOEBA yields the correct trends for the isotopic substitution of H by D with an increase of the heat capacity $c_p$ and of $\Delta H_{vap}$. For the latter, the amplitude of the change is overestimated by a factor of 2 approximately, with a change of $\sim 0.6$~kcal.mol$^{-1}$ for Q-AMOEBA, with respect to the experimental value of 0.3~kcal.mol$^{-1}$. Concerning $c_p$, the isotope effect is also overestimated by Q-AMOEBA (adQTB) with an increase of $\sim$ 3.6~kcal.mol$^{-1}$ compared to the experimental 1.9~kcal.mol$^{-1}$. The Q-AMOEBA (PIMD) model is more accurate in its prediction of the increase of $c_p$ under deuteration but it slightly overestimates the heat capacity in general (for both H$_2$O and D$_2$O). This different behavior of the two Q-AMOEBA variants for this observable might be related to the slight differences in the intramolecular parameters of the two models.
It is very promising that the Q-AMOEBA model is able to capture qualitatively well the effects of the deuteration of water, even though the changes of the properties with isotope substitution are sometimes overestimated (it was also the case for the isotope shift of the temperature of maximum density $T_{MD}$, discussed in section~\ref{sec:thermo_prop}). The remaining discrepancies indicate that, within the tunability offered by the AMOEBA functional form, the optimization via FB leads to parameters that yield slightly too weak hydrogen bonds, which in turns leads to an overestimation of isotope effects since weak H-bonds tend to be further weakened by NQEs~\cite{Ceriotti2016, li_michaelides2011_Hbond}. This interpretation is also consistent with the slight underestimation of $\Delta H_{vap}$ by Q-AMOEBA and with the diffusion coefficient found for D$_2$O: 1.45$\pm 0.02 \times$ $10^5$ cm$^2$.s$^{-1}$ compared to the experimental value of 1.77$ \times$ $10^5$ cm$^2$.s$^{-1}$ \cite{Ceriotti2016,price2000temperature}.
More refined FFs (such as the AMOEBA+ functional form), when suitably re-parametrized to include NQEs explicitly, might, in the future, offer even more quantitatively accurate predictions for isotope effects in water and biological systems. 

\subsection{Ice I$_c$}

The study was further extended by studying the density of ice phase $\text{I}_c$ at 78.0 K under 0~atm of Q-AMOEBA. 128 beads were used to converge PIMD simulations.  
Q-AMEOBA gives 0.893 and 0.906 g.cm$^{-3}$ in PIMD and adQTB respectively whereas the experimental value is 0.931 g.cm$^{-3}$ \cite{ refId0} thus giving a maximum difference of $\sim4\%$. Although classical MD simulations with AMOEBA03 yield a density $\sim2\%$ smaller than the experimental value, PIMD simulations with the same model underestimate the density by $\sim 6.8\%$. Hence, even if Q-AMOEBA probably tends to slightly overemphasize NQEs, it still improves considerably compared to PIMD simulations with AMOEBA03. This indicates that the satisfactory results obtained for Ice with AMOEBA03 in classical MD simulations are due to error compensations and an implicit inclusion of NQEs in the FF parametrization, which limits its transferability. We therefore expect Q-AMOEBA to offer overall more robust predictions in this low-temperature range.

\section{Conclusion}

The newly introduced Q-AMOEBA model allows capturing NQEs with an improved accuracy thanks to their explicit inclusion in the parametrization process. The computational load of this parametrization was significantly reduced by resorting to the adaptive Quantum Thermal Bath method. The final Q-AMOEBA model is significantly different from the original AMOEBA03 force field and the explicit inclusion of NQEs led to an improved intramolecular potential with O-H bond length and H-O-H angle parameters closer to their gas phase values compared to the original AMOEBA03. While Q-AMOEBA is shown to accurately reproduce gas phase quantum chemistry computations, even for large molecular aggregates, it is also shown to be extremely robust for the prediction of condensed phase properties. Improved results on liquid water compared to AMOEBA03 are observed while transferability to ice is shown to be similar to previous models. 
The IR absorption spectrum is also better reproduced with Q-AMOEBA.
 Importantly, the model is also able to capture the experimental trends associated to the deuteration of water. Isotope effects are a valuable indicator of NQEs since they are absent in a classical description of the nuclei. Q-AMOEBA yields qualitatively correct predictions for the changes of thermodynamic quantities in heavy water compared to normal water, although it tends to overestimate these changes. We interpret it as a sign that the hydrogen bonds strength is slightly underestimated by Q-AMOEBA, which should be improved with more advanced functional forms such as AMOEBA+ and improved reference ab initio computations including higher level of couple cluster excitations.  
Concerning the functional form of the force field, it is shown that the inclusion of the many-body polarization effects lead to non-trivial interplay with NQEs. Since adQTB simulations can be performed at near classical cost, the improved Q-AMOEBA model can be easily extended to organic molecules, proteins and nucleic acids, opening the possibility for the large scale study of the importance of NQEs in biophysics.

\section{Technical Appendix}
\subsection{\label{app:fb_deriv}Derivatives of average properties with respect to parameters in the path integrals formalism} 
In order to optimize, the parameters of the force field using experimental data on average properties (for example densities, enthalpy of vaporization, etc...), ForceBalance must be able to compute derivatives of these properties with respect to the FF parameters. While it is possible to simply use finite differences to numerically perform these derivatives, it is well-known that they are very sensitive to statistical noise on average values (computed over a molecular dynamics simulation) so that very long simulations would be necessary. Furthermore, one would have to perform multiple simulations with slightly different parameters in order to compute a single derivative. In ref.~\cite{wang2013systematic}, the authors propose to use the explicit form of the partition function in order to compute parametric derivatives as an average value over a single MD simulation with the current parameters. For a parameter $\lambda$ and an observable $A(\vb{r};\lambda)$, the derivative can be expressed in the classical NPT ensemble as:
\begin{equation}
 \pdv{\expval{A}_\lambda}{\lambda}=\expval{\pdv{A}{\lambda}}_\lambda - \beta\expval{A\pdv{E}{\lambda}}_\lambda -\expval{A}_\lambda \expval{\pdv{E}{\lambda}}_\lambda
\end{equation}
\newline
where $E(\vb{r},V;\lambda)$ is the potential energy and $\expval{\cdot}_\lambda$ denotes an expectation value over the distribution $\rho(\vb{r};\lambda)~\propto~e^{-\beta(E(\vb{r},V;\lambda)~+~PV)}$.
\par
This expression can be generalized to PI simulations using the same process with the path integral partition function. Let us denote $\expval{\cdot}_{N,\lambda}$ the expectation value over the N-beads path integral distribution:
\begin{equation}
    \rho_N(\vb{r_1},\hdots,\vb{r_N};\lambda)=\frac{e^{-\beta\left [ \sum\limits_{i=1}^N \frac{E(\textbf{r$_i$};V;\lambda)}{N}+K(\textbf{r$_1$},\hdots,\textbf{r$_N$})+PV\right]}}{\mathcal{Z}_N(\lambda)}
\end{equation}
with $K(\vb{r_1},\hdots,\vb{r_N})$ the PI harmonic energy that couples the different beads (which is independent of $\lambda$) and $\mathcal{Z}_N(\lambda)$ the partition function that normalizes $\rho_N$. For convenience, let us denote $\overline{O}=\sum_{i=1}^N O(\vb{r_i},V;\lambda)/N$ the average over the beads for any function $O$. 
The parametric derivative of $\expval{A}_{N,\lambda}$ is then:
\begin{align*}
    \frac{\partial \expval{A}_{N,\lambda}}{ \partial \lambda}&=\frac{\partial}{\partial \lambda}\left [\int \text{d}\textbf{r$_1$}\hdots\text{d}\textbf{r$_N$} \overline{A}(\lambda)\rho_N(\textbf{r$_1$},\hdots,\textbf{r$_N$};\lambda)\right]\\
    &=\left \langle \frac{\partial \overline{A}}{\partial \lambda}\right \rangle_{N,\lambda}-\beta\left \langle \overline{A}(\lambda) \left (\frac{1}{N}\sum\limits_{i=1}^{N}\frac{\partial E(\textbf{r$_i$};V;\lambda)}{\partial \lambda}\right)\right\rangle_{N,\lambda}-\langle \overline{A}(\lambda) \rangle_{N,\lambda}\frac{1}{\mathcal{Z}_N(\lambda)}\frac{\partial \mathcal{Z_N}}{\partial \lambda} \\
    &=\left \langle \frac{\overline{\partial A}}{\partial \lambda}\right \rangle_{Z,N}-\beta\left[\left \langle \overline{A}\,\frac{\overline{\partial E}}{\partial \lambda} \right \rangle_{N,\lambda}-\langle \overline{A}\rangle_{N,\lambda}\left\langle\frac{\overline{\partial E}}{\partial \lambda}\right \rangle_{N,\lambda}\right]
    \end{align*}
which is almost identical to the classical expression except that all observables are averaged over the beads. This expression can be used in Force Balance in order to fit the force field parameters on properties computed using quantum nuclei in the path integrals framework.

\subsection{Pressure Estimator}
\label{app:pressure}
NPT simulations were performed using the Langevin-piston method to obtain all relevant constant-pressure properties. The isotropic pressure estimator is given by:
\begin{equation}
\label{eq_pressure}
    P_{int}=\frac{2K}{3V}-\frac{dU}{dV}
\end{equation}
with \textit{V} the volume of the simulated box and \textit{U} the interatomic potential energy. In the PIMD framework, the kinetic energy \textit{K} is given by the centroid virial estimator whereas in the adQTB it is obtained as:
\begin{equation}
    \label{kinetic adqtb}
    K(v_1,v_2,\hdots,v_{3N})=\eta^{-1}\sum_{i}\frac{1}{2}m_iv_i^2
\end{equation}
where $\eta$ is a correction factor obtained through a deconvolution procedure in order to correct for the systematic error in the adQTB estimation of the kinetic energy due to the spectral broadening induced by the Langevin dynamics with relatively large friction coefficients $\gamma$ \cite{doi:10.1021/acs.jpclett.1c01722}. Despite this powerful procedure to correct the pressure estimation in adQTB, some differences remain with respect to PIMD calculations with the same set of parameters. More precisely, within the Q-AMOEBA (PIMD) model and for NVT simulations at 300~K, the average of the second term in equation (\ref{eq_pressure}) is -12840 atm and -12680 atm in adQTB and PIMD respectively, \textit{i.e.} a difference of $\sim1.3\%$. The averages of the term (kinetic part) are equal respectively to 13040 and 12640 atm ($\sim 3.2\%$ difference, though the PIMD value might still increase slightly for larger bead numbers). These discrepancies result in a $\sim 160$~atm difference in the total pressure estimation.
Although relatively small for condensed phase simulations, this discrepancy causes a slight error on the density estimation in adQTB (of the order of 1\%) which is the main reason for the use of two different Q-AMOEBA sets of parameters for adQTB and PIMD. 

\begin{acknowledgement}
This work has received funding from the European Research Council (ERC) under the European Union’s Horizon 2020 research and innovation program (grant agreement No 810367), project EMC2 (JPP). Computations have been performed at GENCI (TGCC, Bruyères le Châtel) on grant no A0070707671. 
\newline

\end{acknowledgement}
\begin{figure}
    \centering
    \includegraphics{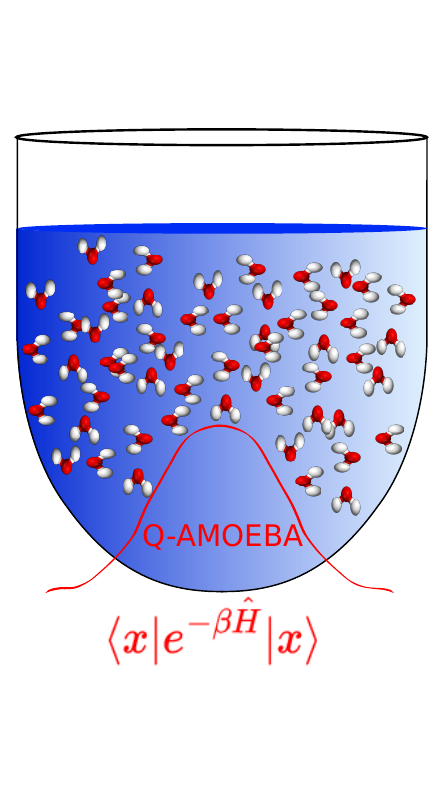}
\end{figure}

\begin{suppinfo}
Supplementary information is available and includes the final parameters of the Q-AMOEBA force field. Details setup of the ForceBalance parameters optimization procedure is also available.
\end{suppinfo}

\normalem
\bibliography{achemso-demo}

\end{document}



\title[]{Improving Condensed Phase Water Dynamics with Explicit Nuclear Quantum  Effects: the Polarizable Q-AMOEBA Force Field}
\author{Nastasia Mauger}
\affiliation{Sorbonne Université, LCT, UMR 7616 CNRS, F-75005, Paris, France}
\author{Thomas Plé}
\affiliation{Sorbonne Université, LCT, UMR 7616 CNRS, F-75005, Paris, France}
\author{Louis Lagardère*}
\affiliation{Sorbonne Université, LCT, UMR 7616 CNRS, F-75005, Paris, France}
\affiliation{CNRS, Sorbonne Université, Institut des NanoSciences de Paris,
UMR 7588, 4 Place Jussieu, F-75005 Paris, France}
\author{Simon Huppert*}
\affiliation{CNRS, Sorbonne Universit\'e, Institut des NanoSciences de Paris,
UMR 7588, 4 Place Jussieu, 75005 Paris, France}
\author{Jean-Philip Piquemal*}
\affiliation{Sorbonne Université, LCT, UMR 7616 CNRS, F-75005, Paris, France}
 \altaffiliation[Also at ]{Institut Universitaire de France, 75005, Paris , France}
 \altaffiliation[Also at ]{Department of Biomedical Engineering, The University of Texas at Austin, USA}
\email{jean-philip.piquemal@sorbonne-université.fr}

\date{\today}
\begin{abstract}
\end{abstract}

\maketitle
\section{ Initial and Optimized Parameters of the Q-AMOEBA water model}
\begin{table}[h]
\centering
\resizebox{\textwidth}{!}{%
\begin{tabular}{cccccc}
\hline \hline
term   & parameter            & unit                       & initial  & Q-AMOEBA (adQTB) & Q-AMOEBA (PIMD)  \TBstrut  \TBstrut  \TBstrut \\ 
\hline \hline
vdW    & O vdW diameter       & \AA                        & 3.4050 & 3.405099         &   3.405099          \TBstrut  \TBstrut    \\
       & O vdW epsilon        & kcal.mol$^{-1}$            & 0.1100 & 0.120139         &   0.120190             \TBstrut   \\
       & H vdW diameter       & \AA                        & 2.6550 & 2.643027         &   2.642967           \TBstrut    \\
       & H vdW epsilon        & kcal.mol$^{-1}$            & 0.0135 & 0.009943      &      0.010287             \TBstrut   \\
       & H vdW reduction      & none                       & 0.910 & 0.93429       &   0.937788          \TBstrut   \\
bonded & O-H bond length      & \AA                        & 0.9572     & 0.9472            &   0.9572            \TBstrut  \\
       & H-O-H angle          & \AA                        & 108.50   & 106.20           &      107.20           \TBstrut  \\
      \hline \hline
      
\end{tabular}%
}
\caption{Value of the initial/AMOEBA03 \cite{ren2003polarizable} water parameters compared to the ones obtained with FB \cite{wang2013systematic,wang2014building} with the adQTB and PIMD methods.}
\vspace{8.9cm}
\end{table}
\hspace{8.0cm}S\thepage
\newpage

\section{Options used for Force Balance}
\begin{Verbatim}
$options
ffdir forcefield
penalty_type L2
tinkerpath  /home/mauger/ForceBalance/ForceBalance_exec/
jobtype newton
forcefield water.prm

maxstep 100

convergence_objective 0.01
convergence_step 0.001
convergence_gradient 0.01
finite_difference_h 0.0001


trust0 0.25
mintrust 0.05
error_tolerance 2.0
adaptive_factor 0.2
adaptive_damping 0.5
normalize_weights no

print_hessian
print_gradient


writechk pimd.chk


amoeba_pol mutual
amoeba_eps 1e-5

constrain_charge false
$end

\end{Verbatim}
\newpage
\section{Q-AMOEBA Force Field Parameters}
\subsection{Q-AMOEBA (adQTB)}
\begin{Verbatim}
forcefield              AMOEBA-WATER

bond-cubic              -2.55
bond-quartic            3.793125
angle-cubic             -0.014
angle-quartic           0.000056
angle-pentic            -0.0000007
angle-sextic            0.000000022
vdwtype                 BUFFERED-14-7
radiusrule              CUBIC-MEAN
radiustype              R-MIN
radiussize              DIAMETER
epsilonrule             HHG
dielectric              1.0
polarization            MUTUAL


      #############################
      ##                         ##
      ##  Literature References  ##
      ##                         ##
      #############################


P. Ren and J. W. Ponder, "A Polarizable Atomic Multipole Water Model
for Molecular Mechanics Simulation", J. Phys. Chem. B, 107, 5933-5947
(2003)

Y. Kong, "Multipole Electrostatic Methods for Protein Modeling with
Reaction Field Treatment", Ph.D. thesis, DBBS Program in Molecular
Biophysics, Washington University, St. Louis, August, 1997  [available
online from http://dasher.wustl.edu/ponder/]

alternative valence parameters to match symmetric and antisymmetric
bond stretches by David Semrouni, Ecole Polytechnique, Paris


      #############################
      ##                         ##
      ##  Atom Type Definitions  ##
      ##                         ##
      #############################


atom          1    1    O     "AMOEBA Water O"               8    15.995    2
atom          2    2    H     "AMOEBA Water H"               1     1.008    1


     ################################
      ##                            ##
      ##  Van der Waals Parameters  ##
      ##                            ##
      ################################


vdw           1               3.405098854695e+00     1.201394077792e-01 
vdw           2               2.643026854641e+00     9.942666570498e-03      9.342808959902e-01 


      ##################################
      ##                              ##
      ##  Bond Stretching Parameters  ##
      ##                              ##
      ##################################


bond          1    2          556.85     0.9472   


      ################################
      ##                            ##
      ##  Angle Bending Parameters  ##
      ##                            ##
      ################################


angle         2    1    2      48.70     106.20 


      ###############################
      ##                           ##
      ##  Urey-Bradley Parameters  ##
      ##                           ##
      ###############################


ureybrad      2    1    2      -7.60     1.5326 


      ###################################
      ##                               ##
      ##  Atomic Multipole Parameters  ##
      ##                               ##
      ###################################


multipole     1   -2   -2              -0.51966                           
                                        0.00000    0.00000    0.14279     
                                        0.37928                           
                                        0.00000   -0.41809                
                                        0.00000    0.00000    0.03881      

multipole     2    1    2               0.25983                           
                                       -0.03859    0.00000   -0.05818    
                                       -0.03673                           
                                        0.00000   -0.10739             
                                       -0.00203    0.00000    0.14412   


      ########################################
      ##                                    ##
      ##  Dipole Polarizability Parameters  ##
      ##                                    ##
      ########################################


polarize      1           0.837      0.390       2     
polarize      2           0.496      0.390       1     

\end{Verbatim}
\newpage
\subsection{Q-AMOEBA (PIMD)}
\begin{Verbatim}
forcefield              AMOEBA-WATER

bond-cubic              -2.55
bond-quartic            3.793125
angle-cubic             -0.014
angle-quartic           0.000056
angle-pentic            -0.0000007
angle-sextic            0.000000022
vdwtype                 BUFFERED-14-7
radiusrule              CUBIC-MEAN
radiustype              R-MIN
radiussize              DIAMETER
epsilonrule             HHG
dielectric              1.0
polarization            MUTUAL


      #############################
      ##                         ##
      ##  Literature References  ##
      ##                         ##
      #############################


P. Ren and J. W. Ponder, "A Polarizable Atomic Multipole Water Model
for Molecular Mechanics Simulation", J. Phys. Chem. B, 107, 5933-5947
(2003)

Y. Kong, "Multipole Electrostatic Methods for Protein Modeling with
Reaction Field Treatment", Ph.D. thesis, DBBS Program in Molecular
Biophysics, Washington University, St. Louis, August, 1997  [available
online from http://dasher.wustl.edu/ponder/]

alternative valence parameters to match symmetric and antisymmetric
bond stretches by David Semrouni, Ecole Polytechnique, Paris


      #############################
      ##                         ##
      ##  Atom Type Definitions  ##
      ##                         ##
      #############################


atom          1    1    O     "AMOEBA Water O"               8    15.995    2
atom          2    2    H     "AMOEBA Water H"               1     1.008    1


     ################################
      ##                            ##
      ##  Van der Waals Parameters  ##
      ##                            ##
      ################################
vdw           1               3.405098077410e+00     1.201901058311e-01 
vdw           2               2.642966815889e+00     1.028645491727e-02      9.377881382906e-01


      ##################################
      ##                              ##
      ##  Bond Stretching Parameters  ##
      ##                              ##
      ##################################


bond          1    2          556.85     0.9572


      ################################
      ##                            ##
      ##  Angle Bending Parameters  ##
      ##                            ##
      ################################


angle         2    1    2      48.70     107.20 


      ###############################
      ##                           ##
      ##  Urey-Bradley Parameters  ##
      ##                           ##
      ###############################


ureybrad      2    1    2      -7.60     1.5337 


      ###################################
      ##                               ##
      ##  Atomic Multipole Parameters  ##
      ##                               ##
      ###################################


multipole     1   -2   -2              -0.51966                           
                                        0.00000    0.00000    0.14279     
                                        0.37928                           
                                        0.00000   -0.41809                
                                        0.00000    0.00000    0.03881      

multipole     2    1    2               0.25983                           
                                       -0.03859    0.00000   -0.05818    
                                       -0.03673                           
                                        0.00000   -0.10739             
                                       -0.00203    0.00000    0.14412   


      ########################################
      ##                                    ##
      ##  Dipole Polarizability Parameters  ##
      ##                                    ##
      ########################################


polarize      1           0.837      0.390       2     
polarize      2           0.496      0.390       1     

\end{Verbatim}



\bibliography{achemso-demo}